\documentclass[12pt,oneside,letterpaper]{report}

\newcommand{\draftfinal}[2]{\ifdefined\draftversion#1\else#2\fi}

\newcommand{\finalonly}[1]{\draftfinal{}{#1}}

\newcommand{\thesistitle}{Theory and Cosmology of \\ Biverse Volume-Normalized Gravity}
\newcommand{\thesisauthor}{Siqing Yu}
\newcommand{\thesisadvisor}{Dr. Gregory Gabadadze}
\newcommand{\thesisdept}{Physics}
\newcommand{\gradmonth}{May}
\newcommand{\gradyear}{2022}

\RequirePackage[margin=1in, includefoot, letterpaper]{geometry}

\RequirePackage[prologue]{xcolor}
\definecolor[named]{ThesisBlue}{cmyk}{1,0.1,0,0.1}
\definecolor[named]{ThesisYellow}{cmyk}{0,0.16,1,0}
\definecolor[named]{ThesisOrange}{cmyk}{0,0.42,1,0.01}
\definecolor[named]{ThesisRed}{cmyk}{0,0.90,0.86,0}
\definecolor[named]{ThesisLightBlue}{cmyk}{0.49,0.01,0,0}
\definecolor[named]{ThesisGreen}{cmyk}{0.20,0,1,0.19}
\definecolor[named]{ThesisPurple}{cmyk}{0.55,1,0,0.15}
\definecolor[named]{ThesisDarkBlue}{cmyk}{1,0.58,0,0.21}

\definecolor{SchoolColor}{rgb}{0.3412, 0.0235, 0.5490} 
\definecolor{chaptergrey}{rgb}{0.2600, 0.0200, 0.4600} 
\definecolor{midgrey}{rgb}{0.4, 0.4, 0.4}

\usepackage{hyperref}
\hypersetup{colorlinks,
  linkcolor=ThesisDarkBlue,
  citecolor=ThesisPurple,
  urlcolor=ThesisDarkBlue,
  filecolor=ThesisDarkBlue}

\RequirePackage[labelfont={bf,sf,small,singlespacing},
                textfont={sf,small,singlespacing},
                margin=0pt,
                figurewithin=chapter,
                tablewithin=chapter]{caption}

\usepackage{fix-cm}
\RequirePackage[raggedright,sc]{titlesec}
\definecolor{gray75}{gray}{0.75}
\newcommand{\hsp}{\hspace{20pt}}

\titleformat{\chapter}[hang]
{\Huge\sc}
{\textcolor{SchoolColor}{\thechapter}\hsp\textcolor{gray75}{|}\hsp}
{0pt}{\Huge\sc\raggedright}

\usepackage{setspace}

\finalonly{
  \doublespacing 
}

\input{defs}

\usepackage{layout} 
\begin{document}

\pagenumbering{roman}
%
\thispagestyle{empty}
%

\vspace*{10pt}
\begin{center}

  {\Large
    \begin{doublespace}
      {\textcolor{SchoolColor}{\textsc{\thesistitle}}}
    \end{doublespace}
  }
  \vspace{.6in}

  by
  \vspace{.6in}

  \thesisauthor
  \vfill

  \begin{doublespace}
    \textsc{
    A dissertation submitted in partial fulfillment\\
    of the requirements for the degree of\\
    Doctor of Philosophy\\
    Department of \thesisdept\\
    New York University\\
    \gradmonth, \gradyear}
  \end{doublespace}
\end{center}
\vfill

\noindent\makebox[\textwidth]{\hfill\makebox[2.5in]{\hrulefill}}\\
\makebox[\textwidth]{\hfill\makebox[2.5in]{\hfill\thesisadvisor}}
\vspace{.6in}

\newpage

\thispagestyle{empty}
\vspace*{25pt}
\begin{center}
  \scshape \noindent \small \copyright \  \small  \thesisauthor \\
  All rights reserved, \gradyear
\end{center}
\vspace*{0in}
\newpage

\setlength{\voffset}{-1in}


\newpage 
\thispagestyle{empty}
\ 

\newpage

\thispagestyle{empty}
\vspace*{\fill}
\begin{center}
  {\it \Large Physics thrives on crisis.}

\bigskip
\bigskip
\bigskip

-- Steven Weinberg, The Cosmological Constant Problem \cite{Weinberg:1988cp}
\end{center}
\vspace*{\fill}

\setlength{\voffset}{-1in}

\chapter*{Acknowledgements}
\addcontentsline{toc}{chapter}{Acknowledgments}
I would like to thank my advisor, Gregory Gabadadze, for his invaluable guidance and unwavering support during my exploration of modern theories of modified gravity and cosmology, as well as his deep insights on the connections between quantum gravity and holographic duality. I would also like to thank my dissertation committee members, Matthew Kleban, Massimo Porrati, Allen Mincer, and Sergei Dubovsky, for their frequent feedback and inspiring input during my thesis work. I am also immensely grateful to my peers, Giorgi Tukhashvili, Daniel Older, Cedric Yu, Xingchen Xu, Di Liu, and Xinyang Yu for their thought-provoking discussions in our shared adventure through the landscape of theoretical physics. I am also indebted to our graduate program administrator, Evette Ma, for her tireless work and support. Last but not least, I would like to thank my parents, Zhenhong Yu and Jianya Qin, who nurtured my interest in physics from a very young age and encouraged my academic pursuit throughout the years.



\chapter*{Abstract}

\setlength{\textheight}{550pt}
\addcontentsline{toc}{chapter}{Abstract}

This thesis expounds the biverse volume-normalized gravity, a novel theoretical framework of classical and quantum gravity that provides an unusual solution to the cosmological constant problem, the 60-degree-of-magnitude discrepancy between the value of the vacuum energy density in our universe predicted by particle physics and that implied by cosmological data. In the literature, there exist two interesting ideas: 1) to use the ``volume-normalized" action functional, the usual action of the particle Standard Model and general relativity but divided by the spacetime volume, first proposed by Tseytlin; 2) to generate the vacuum energy density dynamically by assigning a ghost-free massive potential to the graviton. However, each has a problem: Tseytlin’s attempt suffers from diverging quantum loops and the massive gravity theory has a second metric tensor with questionable origin and the quantum strong coupling problem. Inspired by these two ideas, as well as the AdS/CFT correspondence in string theory, the biverse volume-normalized gravity extends Tseytlin’s idea by placing our (3+1)-dimensional universe on the boundary of a (4+1)-dimensional bulk space. This new adaptation controls the large quantum gravity loops in the limit of vanishing ratio of the spacetime volume of our universe to that of the other universe. The ghost-free massive gravity can then be seamlessly incorporated, not only that its otherwise nebulous second metric tensor appears naturally as the pullback of the bulk metric on the boundary, but its strong coupling problem also solved. The biverse volume-normalized gravity is also consistent with all tested models of particle cosmology and the evolution of the universe, and nontrivially predicts that the tensor-to-scalar ratio of our universe is exactly zero.

\newpage

\setlength{\voffset}{0in}
\setlength{\textheight}{550pt}

\tableofcontents

\newpage

\pagenumbering{arabic} 


\chapter{Introduction} \label{ch1}
Ever since the theoretical development of the particle physics Standard Model (SM) in the 1970s and its subsequent experimental confirmation culminating in the discovery of Higgs boson in 2012, we have acquired a very good understanding of the fundamental physics of the microscopic world up to an energy level of at least 1 TeV, equivalent to a length scale of $10^{-17}$ cm. In the meantime, the advances in general relativity (GR), Big Bang cosmology, and the theory of inflation since the 1980s, as well as the overwhelming experimental evidence supporting them, have equipped us with a tremendous knowledge of the evolutionary history and the large scale structure of the universe we live in, up to a length scale of at least 10 billion light-years, or 10$^{28}$ cm. Just as Weinberg had commented in 1989, physics seems to be running out of crises \cite{Weinberg:1988cp}, and yet it looks more so at the present day.

However, at the junction of these extremities of length scales, there remains an old and fundamental mystery: the cosmological constant (CC) problem. In short, it is the enormous discrepancy between the vacuum energy density predicted by quantum field theory and that implied by the observed cosmological data, with the former larger than the latter by somewhat 60 orders of magnitude. To see this, in the formulation of general relativity, the Einstein field equations for the metric tensor $g_{\mu\nu}$ may be written as
\be
R_{\mu \nu }-\frac{1}{2}g_{\mu \nu}R = \frac{1}{M_{\rm P}^2}T_{\mu\nu} - \lambda_{\rm eff} g_{\mu \nu},
\ee
where $M_{\rm P} = \sqrt{\frac{1}{8\pi G}} \sim 10^{19}$ GeV is the Planck mass, beyond which GR as an effective quantum field theory must be corrected by unknown high energy effects. Here, everything proportional to $g_{\mu \nu}$ is included in the term $\lambda_{\rm eff} g_{\mu \nu}$, so that $\lambda_{\rm eff}$ has contributions from both a free parameter $\lambda$, known as the bare cosmological constant, as well as the vacuum energy density $\langle \rho \rangle$ generated by the quantum fluctuations of all the matter fields. In an illustrative calculation following Weinberg \cite{Weinberg:1988cp}, for a field with mass $m$, we can sum the zero-point energy of all its normal modes up to a momentum cutoff $\Lambda \gg m$. This adds to the total vacuum energy density by
\be
\int_{0}^{\Lambda}\frac{4\pi k^2dk}{(2\pi)^3}\frac{\sqrt{k^2+m^2}}{2} \simeq \frac{\Lambda^4}{16\pi^2}.
\ee
If we consider only the Standard Model fields, then this contribution to the vacuum energy density is expected to be at least (TeV)$^4$. On the other hand, in the standard Big Bang cosmology, the dark energy density is modeled by a cosmological constant and causes an acceleration in the expansion of our universe. With the observed supernova data which shows that our universe has indeed entered a phase of accelerated expansion recently \cite{sn1, sn2}, if we trust GR, then the vacuum energy density must be around $10^{-30}$ g/cm$^3 \sim 10^{-12}$ (eV)$^4$, roughly 60 orders of magnitude smaller than the natural value from the Standard Model. Because $\lambda_{\rm eff}$ receives quantum zero-point energy contribution from every known and unknown matter in physics at all energy scales, if we were to renormalize it using the bare cosmological constant, the total counterterm would have extreme fine-tuning, and the resulting small $\lambda_{\rm eff}$ would be distastefully technical unnatural from a theoretical point of view, as it is not protected by any symmetry in GR, unlike the way small fermion masses are protected by the chiral symmetry in the Standard Model.

To resolve the cosmological constant problem, physicists first go into the direction of finding a technically natural small vacuum energy density. Early approaches try to find new symmetries in supersymmetric field theories, or invoking dynamics analogous to the Peccei-Quinn mechanism for the strong CP problem to adjust the value of the vacuum energy density. Some of these attempts are well reviewed in Weinberg's classic paper \cite{Weinberg:1988cp}. While the former category more often than not leads to more loose ends, the latter can explain for an arbitrary value of the vacuum energy density, but it then still remains to explain why the vacuum energy density is tiny yet nonzero, as would have been implied by the observed rate of accelerated expansion of our universe \cite{sn1, sn2}.

A more precise and inspiring way of stating the cosmological constant problem is to ask why the vacuum energy gravitates so little \cite{Arkani-Hamed:2002ukf}. The answer to this question could be that the vacuum energy density is indeed small for some reason, but it could also be that while the vacuum energy density keeps its natural value, the strength by which it gravitates is much weaker than ordinary matter. This line of thought then leads to models of infrared (IR) modification of gravity, because the cosmological constant problem manifests itself only at very low energy and very large distance scales. In terms of cosmology, a viable IR-modified gravitational theory should in some way allow background metric solutions that manifest accelerated expansion without an explicit cosmological constant term, known as ``self-accelerated" cosmological solutions.

Here we briefly mention two well studied examples of IR modification of gravity, which are conceptually and technically closely related to the main topic of this thesis. The first example is the brane world scenario developed by Dvali, Gabadadze, and Porrati (DGP), where the Standard Model fields are localized on a brane in extra dimensions with infinite volume, so that at large distances, gravity on the brane transitions from being four dimensional to higher dimensional \cite{DGP}. In this limit, the vacuum energy on the brane curves mostly the higher-dimensional bulk, with GR emerging on the brane in its own Einstein-Hilbert action. The induced graviton propagator then becomes highly nonlocal in the IR to make long wavelength sources gravitate weakly, whereas locality is still well preserved for all short wavelength interactions. The cost of nonlocality is then limited to only low energy scales, with the high energy physics effectively left intact. The other example is the Lorentz invariant massive gravity theory developed by de Rham, Gabadadze, and Tolley (dRGT) \cite{deRham:2010ik, deRham:2010kj}. In this theory, the full nonlinear massive graviton self-interaction terms are worked out, such that two of the most pathological problems which generally plague massive gravity theories are cured. First, the van Dam-Veltman-Zakharov (vDVZ) discontinuity \cite{vDVZ1, vDVZ2}, which prevents linearized massive gravity from reducing smoothly to GR in the massless limit because of a residual fifth-force scalar mode, is rendered harmless by the screening effect of the nonlinearities of the scalar itself, a mechanism first proposed by Vainshtein \cite{Arkady}. Second, a sixth degree of freedom that usually accompanies massive gravity, known as the Boulware-Deser ghost \cite{Boulware:1972yco}, is shown to be nonexistent because of the unique structure of the dRGT nonlinear graviton self-interactions \cite{gfree1, gfree2}.\footnote{See also, \cite{gfree3, gfree4, gfree5, gfree6}.} Because the massive graviton self-interactions break the diffeomorphism invariance of GR, a small graviton mass is thus technically natural if the self-interaction structure can be preserved by quantum effects. This makes the dRGT massive gravity a potential candidate for dark energy and thus solution to the cosmological constant problem.

With these two examples in mind, let us start our journey on an unconventional path. Consider the following action functional,
\begin{equation} \label{ac1}
\bar{S}=\frac{S}{V_g}=\frac{\int d^4x~\sqrt{g}\left( \frac{M_{\text P}^2}{2}R+\mathcal{L}_{\rm M} \right)}{\int d^4x~\sqrt{g}},
\end{equation}
where $\mathcal{L}_{\rm M}$ is the Lagrangian containing all matter fields, both known and unknown. With proper regularization of the spacetime integrals, $\bar{S}$ has the property that any vacuum energy $\mathcal{E}$ generated by matter fields simply shift $\bar{S}$ by a contant, $\bar{S} \rightarrow \bar{S} + \mathcal{E}$, and thus does not affect the dynamics of gravity. As such, at the classical level for gravity, the cosmological constant does not curve our universe. Note that this way of getting rid of the CC, namely that CC does not gravitate at all in our universe, is different from unimodular gravity, in which the cosmological constant appears as a constant of integration and is assigned an experimentally determined value. Thus, in the volume-normalized approach, it necessitates some other mechanism of generating the cosmic acceleration in our universe.

This idea of solving the CC problem is first formulated by Tseylin \cite{Tseytlin:1990hn}, based on an earlier idea of Linde \cite{Linde:1988ws} and inspirations 
from the T-duality of string theory. However, as we will see in detail in Chapter \ref{ch2}, this classical solution of the cosmological constant problem can not be trusted once quantum corrections set in, as the graviton loops will have effective Planck constant $\sim \hbar V_g$, which becomes large with big and possibly infinite spacetime volume. Tseytlin's proposal is recently improved by assuming the existence of another universe, nonlocally connected to ours \cite{Gabadadze:2014rwa}. This scenario is described by the action functional
\be
\mathcal{A}=\frac{V_F}{V_g}S+S_F=V_F \bar{S} +S_F,
\ee
where $S_F$ is the Einstein-Hilbert action of the other universe with metric $F_{ab}$ and $V_F$ its diffeomorphism invariant spacetime volume. In Chapter \ref{ch2}, we will see that at the quantum level, in the $g$-universe, matter fields are quantized with $\hbar$ but gravity is quantized with an effective Planck constant $\hbar V_g/V_F$, so that if we can have $V_g/V_F \rightarrow 0$ on classical solutions, we essentially obtain a theory of quantized matter coupled to purely classical gravity in our universe with nonlocal modifications. We will discuss the details of spacetime integral regularization as well as the classical and quantum consistency of $\mathcal{A}$ in Chapter \ref{ch2}. Hereafter we will call $\bar{S}$ the ``volume-normalized" action and the physics derived from it to be in a ``volume-normalized" framework,"  and the full scenario described by $\mathcal{A}$ the ``biverse volume-normalized" action or framework.

With the biverse volume-normalized framework established, we then explore in Chapter \ref{ch3} a few ``more traditional" theories of dark energy and inflation in the volume-normalized framework. The general result is that the volume-normalized framework simply sets the global minimum of any matter potential to zero at the full quantum level. A nontrivial prediction of this is that we have an exactly zero tensor-to-scalar ratio.

In Chapter \ref{ch4}, we start our study on a specific realization of the biverse volume-normalized framework, with our $g$-universe living on the conformal boundary of a bulk AdS$_5$ spacetime. We will discuss how the dRGT massive gravity naturally come up in this setup, when the pull-back of the bulk AdS$_5$ metric finds its place as the fixed fiducial metric of massive gravity on the boundary.

In Chapter \ref{ch5}, we then focus on the gravity solutions in the volume-normalized massive gravity, following the setup in Chapter 4. We present a detailed study of the case where dark energy of our universe is due to dRGT massive gravity. We will calculate the 
new corrections to the the decoupling limit 
Lagrangian and graviton equations of motion in massive gravity adapted to the volume-normalized framework. Just as in vanilla dRGT massive gravity, the new nonlocal terms 
also have internal field space Galilean symmetry, and we call them ``nonlocal Galileons" to make the connection to the Galileon scalar modes in massive gravity. 
We then study a self-accelerated solution and show that the new nonlocal 
terms change the perturbative stability analysis. We will find that the small fluctuations are now stable and nonsuperluminal 
for some simple parameter choices, whereas for the same choices the pure massive gravity 
fluctuations are unstable.  We also present a complete study of the stable spherically symmetric solutions on these 
background solutions.

We finally conclude and discuss the outlook of our theory in Chapter \ref{ch6}.
\\

The new results in this thesis are based on these published works, appearing in order:
i) Gabadadze, G., and Yu, S. (2016). Metamorphosis of the Cosmological Constant and 5D Origin of the Fiducial Metric. {\it Phys. Rev. D}, 94:104059. [arXiv:1510.07943 [hep-th]]. ii) Gabadadze, G., and Yu, S. (2018). Dark energy and inflation with a volume normalized action. {\it Phys. Rev. D}, 98:024047. [arXiv:1611.05833 [hep-th]]. iii) Gabadadze, G., and Yu, S. (2017). Nonlocal Galileons and Self-Acceleration.
{\it Phys. Lett. B}, 768:397-403. [arXiv:1608.01060 [hep-th]]. I was a primary author for all these papers.

\chapter{Revival of an Unconventional Path} \label{ch2}
This chapter is based on the work: Gabadadze, G., and Yu, S. (2016). Metamorphosis of the Cosmological Constant and 5D Origin of the Fiducial Metric. {\it Phys. Rev. D}, 94:104059. [arXiv:1510.07943 [hep-th]].\footnote{Copyright (2016) by the American Physical Society, \url{https://doi.org/10.1103/PhysRevD.94.104059}.}

\section{Tseytlin's Proposal in Equations of Motion}
At the level of equations of motion, we review the mechanism by which a CC is eliminated in Tseytlin's proposal, and look into more detail the role of spacetime nonlocality. Starting with the volume-normalized action (\ref{ac1}), its variation yields
\begin{equation}
\delta \bar{S}\equiv \delta\left( \frac{S}{V_g}\right)=\frac{1}{V_g}\left(\delta S-\frac{S}{V_g}\delta
V_g \right).
\end{equation}
The resulting fields equations can be written as nine traceless equations and one trace equation (with $M_{\rm P}$ set to 1),
\begin{equation}\label{trless}
R_{\mu \nu}-\frac{1}{4}g_{\mu \nu}R=T_{\mu \nu}-\frac{1}{4}g_{\mu \nu}T,
\end{equation}
\begin{equation}\label{tr}
R+T= \langle T \rangle -2 \left\langle g^{\mu \nu} \frac{\partial \mathcal{L}_{\rm M}}{\partial g^{\mu \nu}} \right\rangle, 
\end{equation}
where
\begin{equation}
T_{\mu \nu}=g_{\mu \nu}\mathcal{L}_{\rm M}-2\frac{\partial \mathcal{L}_{\rm M}}{\partial g^{\mu \nu}},
\end{equation}
and $\langle \cdots \rangle$  denotes the spacetime average,
\begin{equation}
\langle \cdots \rangle\equiv\frac{\int d^4x \sqrt{|g|} (\cdots)}{V_g}.
\end{equation}
The traceless equations (\ref{trless}) are exactly the same as those obtained by using the Einstein-Hilbert action, whereas nonlocality enters only in the trace equation (\ref{tr}), comparing to the Einstein-Hilbert one, $R+T=0$.

This modification is dramatic, as the LHS of (\ref{tr}), which are local observables, are then affected by spacetime averaged quantities. These averages, when nonzero, have information of the entire space and also preknowledge of the future, thus formally nonlocal and acausal. However, if $V_g \rightarrow \infty$, then for any observable that is localized in either space or time, its spacetime average is identically zero due to the volume factor suppression. As such, the acausality of the trace equation does not manifest itself in the dynamics of most stuff in the universe. Meanwhile, for a constant term in the Lagrangian, which only affects the trace equation, it contributes to both sides of (\ref{tr}) by the same amount. Therefore, the main consequence of the acausality might just be that we do not observe the big CC the in our universe. The fact that a big CC does not affect local dynamics is more straightforward to see at the level of action functional (\ref{ac1}), $\bar{S}=\frac{S}{V_g}$: a shift of $\mathcal{L}_{\rm M}$ by any constant changes  
$\bar S$ by an additive constant, and the latter does not affect the 
equations of motion obtained  by varying $\bar S$.  

While the volume-normalized proposal appears to be an efficient way to get rid of an arbitrary CC irrespective of its origin, the equations 
of motion obtained  from $\bar S$ reveal the high cost of the proposed solution: in the 
equations of motion, the local Ricci curvatures are determined by 
spacetime averages over past and future volumes, when these 
volumes are well-defined \cite {Tseytlin:1990hn}. This seems to be  
a dramatic departure away from the conventional local field theory paradigm.

Such spacetime nonlocality (and therefore,  acausality \cite{Arkani-Hamed:2002ukf}) is however operative only for a
cosmological constant, while  all the local interactions that  can be  measured in a laboratory -- even if 
in one with the size of the observed universe -- are not affected. 
One may wonder how the mechanism could distinguish  between a cosmological constant
and, say,  a very flat scalar potential.  It does so by being nonlocal in time, and hence being   able 
to target and eliminate  only the cosmic fluid that  does not red-shift in the infinite future \cite{Arkani-Hamed:2002ukf}.  
As a result, a CC is eliminated,  while 
the slow roll inflation, radiation,  and matter dominated epochs of cosmology remain intact.\footnote{The mechanism is applicable to a single vacuum universe  \cite{Tseytlin:1990hn,Linde:1988ws}. Its generalization to the case that adopts chaotic inflation with multiple vacua  
is not known to the authors. Throughout the paper we focus on a scenario  in which our 
4D universe evolves in a single vacuum state.} 

For the noncosmological solutions of Einstein's gravity, since the spatial 
nonlocality of the proposed equations is 
operative only for sources that  have an infinite spatial support, all spatially local solutions of Einstein's gravity remain intact also. Thus, the high cost  
incurred by  the solution of the CC problem, while somewhat bothersome, remains  undemanding,   
at least until inconsistencies of the proposal are found.

We add  that the  action   (\ref {ac1})   can readily be used to obtain the 
classical Hamiltonian of the theory. The multiplier, 
$1/V_g$,   does not contain  any  derivatives of the metric, and 
affects the determination of the  canonical momenta only by rescaling them with 
a multiplicative numerical factor of $1/V_g$. To summarize,  in the context of classical gravity and conventionally 
quantized particle physics,  if  $\bar S$ is used as a gravitational action for a single  
vacuum state in which the universe is presumed to evolve,   
then the CC is removed,  without any  apparent contradiction 
with  observations  (see also \cite{Davidson},  for further 
independent work on the proposal).   

\section{Quantization Procedure and Problem of Tseytlin's Volume-Normalized Action}

Before we further the study of the implications of the volume-normalized framework, we must check whether the volume-normalized solution to the CC problem remains consistent when corrected by quantum effects. As already mentioned by Tseytlin, the graviton loop corrections should be problematic, because they already strongly renormalize the form of the action function (\ref{ac1}) at low energy scales, before we even ask
the question of the existence of a full-fledged UV complete quantum theory of gravity, from which 
the action (\ref {ac1}) might originate.

To see the problem and motivate our biverse volume-normalized extension, we here lay out the rules of quantization for the action functional (\ref{ac1}). The quantization procedure is also unconventional, as we will need two different path integrals, one for all the nongravitational interactions quantized with the Planck constant $\hbar$, and the other for gravity with a different and dynamically determined Planck constant.

In the absence of dynamical gravity, namely the $M_{\rm P} \rightarrow \infty$ limit, $\bar{S}$ differs from the standard action by an overall $1/\infty$ factor, which is field independent and then can by scaled away. In this limit, on would quantize the nongravitational fields in the well developed quantum field theory formalism. However, when dynamical gravity is involved, we need to specify the rules of quantization. We will not immediately worry about a UV completion, but a low energy effect field theory quantization of gravity is certainly a serious concern, because GR is a good low energy effective field theory, and any of its modification should strive to retain this virtue below some energy scale that is lower than the Planck scale. Our goal in this chapter is to define such a theory.

We assume that gravity is quantized at some energy scale, $M_{\rm QG}$, which could be the Planck scale or some string scale, that is at least one order of magnitude higher than the UV scale of all nongravitational interactions, $M_{\rm matter}$. $M_{\rm matter}$ could be a scale at which the Standard Model interactions themselves become UV complete, for instance, by being grand unified into an asymptotically free quantum theory. In such a setup, it is not unnatural to have a hierarchy of some two orders of magnitude between $M_{\rm QG}$ and $M_{\rm matter}$. If so, then gravity should be well approximated by a classical field theory below $M_{\rm matter}$. At these energy levels, which we will refer to as ``low energies" in the following, the path integral can be defined with all the matter fields quantized using $\hbar$, while treating gravity as an external classical field, pending specification of the rules of quantization for gravity, which must give rise to further small corrections to the already quantized nongravitational interactions.

These arguments lead to the path integral for nongravitational fields at low energies,
\be
Z(g, J_n) = {\rm const}~ \int d\mu(\tilde{\psi}_n) \exp \left[ {i\int d^4x\sqrt{g}\left( \mathcal{L}(g, \tilde{\psi}_n) + J_n\tilde{\psi}_n \right)} \right],
\ee
where $d\mu(\tilde{\psi}_n)$ is a measure for all the matter fields that appropriately modes out gauge equivalent classes. The metric field $g$ is an external field, and so are the sources, $J_n$, introduced for every single matter field. Then, following the standard Legendre transform quantization procedure, we can define the effective action $\Gamma_{\rm M}$ and Lagrangian $\mathcal{L}_{\rm M, eff}$ by
\begin{equation}
    \Gamma_{\rm M} = \int d^4x \sqrt{g}\mathcal{L}_{\rm M, eff}(g, \psi_n) = -i\ln Z(g, J_n) - \int d^4x \sqrt{g}J_n\psi_n,
\end{equation}
where $\psi_n = -\frac{i}{\sqrt{g}}\frac{\delta \ln Z(g, J_n)}{\delta J_n}$, namely $\sqrt{g}$ times the vacuum expectation value of the matter field $\tilde{\psi}_n$ in the presence of the source $J_n$. $\Gamma_{\rm M}$ defined as such is a 1PI effective action and all the quantum corrections due to nongravitational fields are taken into account in the effective Lagrangian $\mathcal{L}_{\rm M, eff}$. It is this effective Lagrangian that is inserted into (\ref{ac1}) to account for dynamical gravity. At this point, the effective Lagrangian $\mathcal{L}_{\rm M, eff}(g, \psi_n)$ depends on the classical fields, $g$ and $\psi_n$, only; the difference between these two sets of fields is that the quantum corrections due to the nongravitational interactions are already accounted for, whereas the gravity quantum loops are not considered yet. We also define an effective generating functional
\begin{equation}\label{Zm}
    Z_{\rm M}(g, \psi_n) = e^{i\Gamma_{\rm M}} = Z(g, J_n)\exp \left( {-i\int d^4x \sqrt{g}J_n\psi_n} \right),
\end{equation}
that includes all the nongravitational quantum effects, but does not include quantized gravity.

After this, gravity in $\bar{S}$ needs also to be quantized. The corresponding quantum effects are likely to become order one at scales $M_{\rm QG}$, and then taken care of by a putative UV completion of the theory, presumably via new degrees of freedom that could appear at energies $\sim M_{\rm QG}$. These considerations can be postponed for a UV complete theory of gravity, such as string theory, possibly along the lines proposed by Tseytlin in \cite{Tseytlin:1990hn}. However, there is an immediate issue irrespective of the form of UV completion. It concerns the low energy effective theory: the quantum gravity corrections should be small at scales well below $M_{\rm QG}$ for our volume-normalized approximations to be meaningful. For instance, in Einstein's gravity supplied with a diffeomorphism invariant UV cutoff for gravity loops (which requires additional counterterms to retain diffeomorphism invariance), one generates higher dimensional operators that make negligible contributions at energies below $M_{\rm QG}$. In the present case, however, one first needs to define the rules of calculation for the quantum gravity loops, because the classical action $\bar{S}$ has an unusual form. We will define these rules and check whether gravity loop corrections are small in the remainder of this section. We define the extended action,
\begin{equation} \label{sq}
    \bar{S}_{q,\lambda}=\frac{1}{q}\int d^4x \sqrt{g}\left(\frac{R}{2}+\mathcal{L}_{\rm M, eff}  \right)+\lambda(V_g-q),
\end{equation}
and write down the path integral for gravity as
\be \label{zg1}
Z_g = {\rm const}~\int  d\mu(g) ~\! dq ~\! d\lambda ~ \exp \left( {i\bar{S}_{q,\lambda}} \right),
\ee
where $d\mu(g)$ is a measure over diffeomorphism inequivalent metric fields. In this path integral, the fields of the 1PI matter action, $\psi_n$, play the role of the external fields. One also integrates over real parameters $q$ and $\lambda$, which are formally Lagrange multipliers. Using (\ref{Zm}), $Z_g$ can also be written more explicitly,
\begin{equation} \label{zg2}
    Z_g = {\rm const}  ~\int d\mu(g) ~\! dq ~\! d\lambda~ \exp \left[ \frac{i}{q} \left( S_{\rm EH} + \Gamma_{\rm M} \right) \right] e^{i\lambda(V_g-q)},
\end{equation}
where $S_{\rm EH}$ is the Einstein-Hilbert action for gravity.

The two path integrals $Z_{\rm M}$ and $Z_g$ then define a complete algorithm for calculating quantum corrections. First, the nongravitational quantum effects are calculated in the conventional way, assuming the metric to be an external classical field. Then, to calculate gravity loops, the unconventional prescription in the form of (\ref{zg1}) or (\ref{zg2}) is used, after obtaining the quantum effective action for the matter fields. The form of (\ref{zg2}) suggests that the parameter $q$ may be regarded as a second Planck constant that governs the gravity loops, after the matter field quantum effects being calculated using $\hbar$. As such, when one integrates with respect to $q$, the second Planck constant, and $\lambda$, whose equation of motion constrains $q$ to be the value of the invariant four-volume, one finds (\ref{zg2}) become
\be \label{zg3}
Z_g = {\rm const}~\int d\mu(g)~ e^{i\bar{S}},
\ee
with $\bar{S}$ defined in (\ref{ac1}). The form of the extended action (\ref{sq}), unlike $\bar{S}$, is useful for thinking of the formulation of the path integral, or canonical momenta and Hamiltonian of the theory.\footnote{The idea of integration with respect to the parameters is adopted from \cite{Tseytlin:1990hn}, although the path integral here and its interpretation differ somewhat from that in \cite{Tseytlin:1990hn}.}

The trouble with the gravity loops in the effective field theory approach presented above can be understood in the language of either (\ref{zg2}) or (\ref{zg3}). It is easier to see from (\ref{zg3}) that the $1/V_g$ factor in $\bar{S}$ is rescaling what would have been the Planck constant for the gravity loops in a conventional effective field theory approach to Einstein's gravity, namely, we should take all the quantum gravity loop corrections calculated in the conventional approach and make the replacement $\hbar \rightarrow \hbar V_g$. Adopting this procedure for the gravity loops, one would get the renormalized expansion,
\begin{equation} \label{Sren}
\bar{S}_{\rm ren} = \frac{1}{V_g}\int d^4x \sqrt{g} \left( \frac{R}{2} + \mathcal{L}_{\rm M, eff}(g, \psi_n) + V_g\mathcal{L}_1(g, \psi_n) + V_g^2\mathcal{L}_2(g, \psi_n) + \cdots \right),
\end{equation}
where $\mathcal{L}_1, \mathcal{L}_2, ...$ contain all possible terms consistent with diffeomorphism invariance and internal symmetries of the matter fields. When $V_g$ is huge, these new terms resulting from gravity loops are huge as well, and then ruin the classical solution to the cosmological constant problem.

It should be noted that there is yet another class of loop corrections if one quantizes graviton fluctuations in $\bar{S}$ on a given background  solution of the metric.\footnote{This point is first brought about by Arkady Tseytlin.} Suppose we can decompose the metric into background and fluctuation, $g = g_b + h$ schematically, where $h$ is small. Then the inverse volume $1/V_g$ can be expanded as $V_g^{-1} = V_b^{-1}-V_b^{-2}H_h + \cdots$, where $V_b = \int d^4x \sqrt{g_b}$ and $H_h = \int d^4x \sqrt{g_b}{\rm tr}(g_b^{-1}h)/2$, and so on. It is clear that the term, $-V_b^{-2}H_h$, and all the other subsequent terms containing higher powers of $h$, will produce new unconventional interaction vertices when Wick-contracted with powers of $h$ in the expansion of the action $S$. While an extra effort would be required to work out all these unconventional vertices, all of them will be suppressed by powers of the background volume $V_b$. For example, in $-V_b^{-2}H_h$, one power of the inverse volume $V_b^{-1}$ will be spent to offset the volume factor in the expression $H_h$, while the second power of $V_b^{-1}$, not cancelled, will suppress the fluctuation $h$. Thus, all the new vertices will come suppressed by as many powers of $V_b^{-1}$ as the power of the fluctuation $h$ arising from the expansion of $V_g^{-1}$ in $\bar{S}$. This suggests that the loop effects discussed in the present paragraph could be assumed small and be neglected. Similar considerations are applicable to the biverse volume-normalized scenarios.

\section{The Big Constant Out, the Small Constant In}
In \cite{Gabadadze:2014rwa}, a biverse extension is proposed to avoid the difficulty with quantum gravity loops ruining the classical solution to the CC problem in volume-normalized setup. It is described by the biverse volume-normalized action functional
 \begin{equation}\label{acmain}
\mathcal{A}={V_f}{\bar S}+S_f,
\end{equation}
where $f$ is the metric of a second universe that interacts with our universe $g$ only globally. Following \cite{Gabadadze:2014rwa}, here we use the lower case $f$ to denote the metric of a generic second universe. ${\bar S}$ is defined in (\ref{ac1}), and is a functional of the metric $g$ 
 and all the matter fields therein.  The action $S_f$    
contains the Einstein-Hilbert action for $f$,  but is independent 
of the metric of our universe,  $g$.   From the point of view of the $f$-universe, 
the term   ${V_f}{\bar S}$ represents  an addition to  a CC in  that universe. Following the argument in Section 2.1, it is then straightforward to see  from (\ref {acmain}) that an arbitrary cosmological constant introduced in $S$, the numerator in (\ref{ac1}),
becomes a cosmological constant of the $f$-universe. Therefore, there is no big CC problem in the 
$g$-universe -- the CC of the $g$-universe gets entirely absorbed into the 
$f$-universe, which could be highly curved. However, if the classical solution of $f$ has a curvature length scale much shorter than the one associated with the CC of $g$, then the correction to the CC of $f$ will be a small one. Meanwhile, from the $g$-universe point of view, its quantum gravity loops will then have effective Planck constant $\hbar \frac{V_g}{V_f}$, which could be very small if $V_f \gg V_g$, or zero if $V_g/V_f \rightarrow 0$, thus solving the uncontrolled quantum gravity loop problem in $g$-universe. The goal of this section is then to establish the classical and quantum consistency of this picture.

In the last section, we presented an unconventional procedure of using two path integrals to establish the quantum consistency of the volume-normalized picture, assuming an reasonable hierarchy between the quantum gravity and nongravitational fields scales. We need to do the same for the biverse scenario (\ref{acmain}). To proceed, we write the action for $f$-universe as
\be \label{Sf}
S_f = \int d^Dy \sqrt{f} \left( \frac{M_f^{D-2}}{2}R(y) + c_0M^D + \cdots \right),
\ee
with $D \geq 4$, and $f_{AB}(y)$ with $A, B = 0,1, ... D-1$. The action of the $f$-universe has a certain vacuum energy scale $M$, and a scale that determines the strength of its gravitational coupling, $M_f$. Depending on the details of the theory, encoded in the dots in (\ref{Sf}), there may or may not be a stable hierarchy between the scales $M_f$ and $M$. We will explain this further below.

An analog of the extended action (\ref{sq}) now takes the form,
\begin{equation}
    A_{q, \lambda} = \frac{1}{q}\int d^4x \sqrt{g}\left(\frac{R}{2}+\mathcal{L}_{\rm M, eff}  \right)+\lambda \left( \frac{V_g}{V_f}-q \right) + S_f,
\end{equation}
which can be used to define a path integral that includes integration with respect to $q$ and $\lambda$. As discussed in the last section, this path integral specifies the rules of quantizing $g$ and $f$, endowing the $g$-gravity quantum loops with rescaled Planck constant $\hbar \frac{V_g}{V_f}$, while the $f$-gravity is still quantized with $\hbar$. Analogous to (\ref{Sren}), after integrating out $q$ and $\lambda$, the action including the the gravity loop corrections would take the form
\begin{equation}\label{Sren2}
\begin{split}
    & \frac{V_f}{V_g}\int d^4x\sqrt{g} \left( \frac{R(g)}{2} \mathcal{L}_{\rm M, eff} +  \frac{V_g}{V_f} \mathcal{L}_1 + \cdots \right)  \\
    +& \int d^Dy \sqrt{f} \left( \frac{M_f^{D-2}}{2}R(f) + c_0M^D +c_1R(f)^2 + c_2\frac{\bar{S}^2}{M_f^D} + \cdots  \right).
\end{split}
\end{equation}
As such, as long as $V_f \geq V_g$, all the corrections proportional to $V_g/V_f$ can be safely neglected. There are also terms similar to the ones discussed at the end of the previous section, but they are harmless for the reasons presented there. Thus, the gravity loop corrections in $g$-universe to either the gravity itself or the nongravitational processes through the backreaction on $g$ are small or vanishing in this prescription, although the theory still needs a UV completion to make sense of the unusual path integral for $g$ and $f$ gravities. Meanwhile, the gravity loop diagrams in the $f$-universe generate two groups of new terms, one containing higher powers and derivatives of $R(f)$ and the other containing powers of $\bar{S}$, as well as a mixing of the two. Some of these terms are displayed in (\ref{Sren2}). These terms must be made small, and we will present such an argument invoking the hierarchy between the scales in the $g$- and $f$-universes below, the same one making $V_f/V_g \gg 1$.

To make $V_f/V_g \gg 1$, we explore the technically natural hierarchies between scales of the theory. First, we assume that the $g$-universe has supersymmetry broken at some high scale, and therefore, there is a natural value of its vaccum energy density proportional to $E_{\rm vac}^4$. The scale $E_{\rm vac}$ can be anywhere between TeV and the GUT scale, $\mu_{\rm GUT} \sim 10^{16}$GeV. For the $f$-universe, it is presumably uncontroversial to set $M_f \sim M_{\rm P}$, but also we would need to scale $M$ to be somewhat higher than $E_{\rm vac}$. The latter condition should be natural, since without special arrangements one would expect $M \sim M_f \sim M_{\rm P}$, and since $E_{\rm vac} \ll M_{\rm P}$, one would also get $E_{\rm vac} < M$. If so, then the vacuum energy of the $g$-universe would make a small contribution to the pre-existing vacuum energy of the $f$-universe. In other words, the vacuum energy density of the $f$-universe, $c_0M^D$, would dominate over the vacuum energy density that gets delegated to the $f$-universe from the $g$-universe.

For a concrete $f$-universe picture, while one could try to explore a case when the $f$-universe has a positive vacuum energy density, it is more straightforward to make a mild assumption that the curvature due to the term $c_0M^D$ in the $f$-universe is negative, thus Anti-de Sitter (AdS) like. An intuitive scenario is that the $f$-universe is exactly supersymmetric and described by some unbroken supergravity \cite{Gabadadze:2014rwa}. The work \cite{Gabadadze:2014rwa} has considered a case where $D=4$ and $f$-metric describes $\mathcal{N}=1$ AdS supergravity. The total vacuum energy in $f$-universe then is $3\bar{\lambda} = \bar{S} + c_0M^4$, so that if we can have $M_f \gsim 10\sqrt{\bar{\lambda}}$, or equivalently, $M_{\rm P} \sim M_f \sim 10M$, $M \gsim 10E_{\rm vac}$, then the $f$-universe AdS curvature receives only a small correction for the CC of $g$-universe. In particular, it is noted that this hierarchy can be generated by the Adler-Zee mechanism, so it could be technically natural \cite{Gabadadze:2014rwa}.

To look at the spacetime volume ratio $V_f/V_g$, it is convenient to invoke the Euclidean volumes. For a general $D$-dimensional AdS $f$-universe with $D \geq 4$, $V_f = \infty$ even in Euclidean space, whereas $V_g = H_0^{-4}$ is finite for a dS $g$-universe, if we can have a way to generate a curvature with scale $H_0$ in the $g$-universe other than the CC. The fact that $V_{f, {\rm AdS}}/V_{g, {\rm dS}} \rightarrow \infty$ can also be seen in the Lorentzian signature, by calculating the ratio in the global coordinate systems for the universal covering of AdS and dS spaces. To achieve a small dS curvature in $g$-universe, it can only be due to the terms in $S$ that explicitly depend on $g$, as any CC-like term generated by, say a scalar potential, contributes only to the curvature of $f$-universe. One such possibility is massive gravity in $g$-universe. We will look at this possibility in detail in Section 4 and 5. Before that, we will explain how standard cosmology is modified in the $g$-universe in Section 3.

\chapter{Volume-Normalized Cosmology} \label{ch3}
This chapter is based on the work: Gabadadze, G., and Yu, S. (2018). Dark energy and inflation with a volume normalized action. {\it Phys. Rev. D}, 98:024047. [arXiv:1611.05833 [hep-th]].\footnote{Published under CC BY 4.0, \url{https://doi.org/10.1103/PhysRevD.98.024047}.}

\section{Cosmology Largely Preserved Except the Cosmological Constant}
As is discussed above, the volume-normalized action,
\be \label{ac2}
\bar{S}=\frac{S}{V_g},
\ee
has the property that any cosmological constant generated in the 4D $g$-universe by any source does not curve the spacetime geometry of the $g$-universe itself. When embedded into the extended framework described by the biverse volume-normalized action,
\be \label{pr1}
\mathcal{A}=\frac{V_F}{V_g}S+S_F=V_F \bar{S} +S_F,
\ee
this cosmological constant gets absorbed into $S_F$, correcting the curvature of the bulk space described by the $F$-metric by a small amount, if a hierarchy can be established for the Planck scales of the $g$- and $F$-universes. As a concrete example, the bulk space can be taken as AdS$_5$ space, and the $g$-universe which describes our Universe resides on its conformal boundary. From here on, we use the capital $F$ to accentuate our choice of a higher dimensional second universe.

The full biverse volume-normalized action $\mathcal{A}$ has been introduced for its theoretical advantage of producing a consistent low energy theory of quantum gravity in the $g$-universe. As noted by Tseytlin \cite {Tseytlin:1990hn}, with just $\bar{S}$, the $g$-metric gravity loops should be calculated with an effective Planck constant $\hbar V_g$, leading to large effects of quantum gravity which are expected to ruin the classical solution that eliminates the cosmological constant, at the very least. On the contrary, as explained in Chapter \ref{ch2}, when $\bar{S}$ is embedded in $\mathcal{A}$, the particle physics quantum effects can be kept intact, while the gravity 
loops of $g$-metric are suppressed by a different effective Planck constant, 
$\hbar V_g/V_F$. The ratio of the volumes is determined by the classical solutions such that $\hbar V_g/V_F\to 0$. Then the gravity loops do not generate any counterterms  to the action for the $g$-metric, and 
at low energies, gravity in our Universe becomes a purely classical theory that couples to quantized particle 
physics \cite{Gabadadze:2014rwa,ggsy}. Therefore, when we talk about classical cosmologies in the $g$-universe, we can safely use $\bar{S}$. It is only when quantum effects are in concern then we need to refer to the full biverse volume-normalized action.

It is then essential to introduce a dynamical dark energy component in the $g$-universe in order to explain the accelerated expansion of our universe today. To this regard, a promising candidate utilizes the massive gravity theory proposed by de Rham, Gabadadze, and Tolley (dRGT) \cite{deRham:2010ik,deRham:2010kj}, a theory that has a technically natural graviton mass \cite{dRGHP}. Being ghost-free, it also presents a new class of healthy self-accelerated background solutions in the decoupling limit. With the help of nonlocal Galileon terms \cite{galileon}, these solutions exhibit no linear scalar-tensor mixing \cite {dRGHP_sa} and thus no vDVZ discontinuity, which is common in models of IR modification of gravity \cite{mgreview1}.  However, just as any other solution obtained via the decouplimg limit, their full-theory consistency beyond that limit remains unsettled. We will explore massive gravity with the volume-normalized action in Chapters \ref{ch4} and \ref{ch5}. Reviews of massive gravity can be found in \cite{mgreview1,mgreview2,mgreview3}.

Instead, we first investigate a few ``more traditional" candidate models of dark energy in the volume-normalized framework. We limit ourselves to the following case studies described by the rolling of a single homogeneous scalar field: (a) minimally coupled to gravity, commonly known as ``quintessence" models \cite{quint11,quint22}, (b) nonminimally coupled to gravity \cite{nmcintro,nmcintro1,nmcintro2}, and also (c) Chaplygin gas \cite{chap}. More generally, many of the single scalar field scenarios with a slow-roll regime were first used to describe inflation, to which the results of the following discussions apply equally well, once the energy scales involved are adjusted accordingly \cite{inf,inf2,inf3}.\footnote{Throughout this chapter, we consider only the scenarios in which our 4D Universe evolves in
a single vacuum state. We do not attempt, for instance, to generalize (\ref{ac2}) to adopt chaotic inflation
with multiple vacua.} As such, we can conveniently treat the dark energy and the inflationary scenarios driven by the same mechanism together, addressing them only separately when appreciable differences arise. It turns out that cases (a) and (b) remain unscathed when incorporated into (\ref{ac2}), whereas case (c) loses its original features, due to the fact that the volume-normalized action shifts the global minimum of any scalar potential to zero. We explain the details in the following Sections, \ref{mcs}, \ref{nmcs}, and \ref{cg}, respectively. We then highlight the testable prediction of a zero tensor-to-scalar ratio in the primordial power spectra as a consequence of the volume-normalized framework, after a discussion on the classical and quantum field fluctuations in section \ref{conc}. Reviews can be found in \cite{dereview1,dereview2,dereview3} for dark energy and \cite{infreview1,infreview3,infreview4,infreview5} for inflation.

To study cosmology, we use a homogeneous scalar profile, $\phi(\vec{x},t)=\phi(t)$, on a flat Friedmann-Robertson-Walker (FRW) background metric in the mostly positive signature, $ds^2=-dt^2+a(t)^2\delta_{ij}dx^idx^j$. Time derivatives are denoted by dots. To make nonlocal expressions unambiguous, a ratio between spacetime integrals is defined as follows: first compute the ratio of integrals in a finite spacetime volume, then take the limit when the region of integration tend to infinity in all dimensions, as discussed in the last chapter.

It is noteworthy that the works \cite{Davidson, Davidson2} by Davidson and Rubin have discussed extensively the classical FRW cosmologies allowed by the effective action (\ref{ac2}). Using predominantly minisuperspace techniques, they have showed the vanishing of a rigid cosmological constant in quintessential models \cite{Davidson} and proved that a perfect fluid can only lead to nonclosed universes and a zero cosmological constant \cite{Davidson2}, except for the exotic cases of closed static Einstein universe and Eddington-Lema\^itre universe \cite{el1,el2}. As such, we acknowledge that the classical analysis in Sections \ref{mcs}, \ref{nmcs}, and \ref{cg} have significant conceptual overlaps with \cite{Davidson, Davidson2}. Here, our major difference from \cite{Davidson, Davidson2} is to establish the consistent low energy theory of quantum gravity through the full action (\ref{pr1}), a problem left unaddressed in \cite{Davidson, Davidson2}.
\\

\section{Minimally Coupled Scalar}\label{mcs}
In the framework of (\ref{ac2}), the theory of a scalar field minimally coupled to gravity is described by the variation of
\begin{equation}
\bar{S}=\frac{S}{V_g}=\frac{\int d^4x~\sqrt{g}\left[\frac{M_{\text P}^2}{2}R-\frac{1}{2}g^{\mu \nu}\partial_{\mu}\phi\partial_{\nu}\phi -V(\phi) \right]}{\int d^4x~\sqrt{g}}.
\end{equation}
This modifies the Einstein field equations to
\begin{equation}
M_{\rm P}^2G_{\mu \nu}=T_{\mu \nu}-\bar{S}g_{\mu \nu},
\end{equation}
with
\begin{equation}\label{Tmunu}
T_{\mu \nu}=\partial_{\mu}\phi \partial_{\nu}\phi -g_{\mu \nu}\left[ \frac{1}{2}g^{\alpha \beta}\partial_{\alpha}\phi\partial_{\beta}\phi +V(\phi)   \right] .
\end{equation}
The Friedmann equations become
\begin{gather}
3M_{\text P}^2 H^2=\rho_{\phi}+\bar{S}, \label{f1} \\
3M_{\text P}^2(\dot{H}+H^2)=-\frac{\rho_{\phi}+3p_{\phi}}{2}+\bar{S} \label{f2},
\end{gather}
with the definitions
\begin{equation}
\rho_{\phi}=\frac{1}{2}\dot{\phi}^2+V(\phi),\quad p_{\phi}=\frac{1}{2}\dot{\phi}^2-V(\phi),\quad H=\frac{\dot{a}}{a}, \quad w_{\phi}=\frac{p_{\phi}}{\rho_{\phi}}.
\end{equation}
The $\phi$ equation of motion is still
\begin{equation}
\ddot{\phi}+3H\dot{\phi}+V'(\phi)=0.
\end{equation}

Compared to the familiar theory described by just $S$, there are now new terms $\propto \bar{S}$ in the Friedmann equations. However, using (\ref{f1}) and (\ref{f2}),
\begin{equation}
\bar{S}=\frac{\int d^3xdt~a^3\left[3M_{\text P}^2(\dot{H}+2H^2)+p_{\phi} \right]}{\int d^3xdt~a^3}=\frac{\int d^3xdt~a^3\left[ 2 \bar{S}+\frac{\rho_{\phi}-p_{\phi}}{2} \right]}{\int d^3xdt~a^3},
\end{equation}
so
\begin{equation}\label{sbar1}
\bar{S}=-\frac{\int dt~a^3V(\phi(t))}{\int dt~a^3}=-\lim_{T\rightarrow \infty} \frac{\int_{t_i}^T dt~a^3 \left[ V(\phi(t)) -V_{\min}\right] }{\int_{t_i}^T dt~a^3}-V_{\min},
\end{equation}
where $t_i$ is the time when $\phi$ becomes dynamical and $V_{\min}$ is the global minimum of the potential $V(\phi)$ that $\phi$ rolls towards. Here, we write $\bar{S}$ in such a way that the integrand of the numerator in (\ref{sbar1}) is zero at $t=T$.

It is not hard to see that the first term in $\bar{S}$ written this way vanishes on reasonable potentials and rolling $\phi$ profiles that satisfy the minimal requirements of smoothness, increasing scaling factor and decaying energy density. First, note that $\int dt~a^3V(\phi)$ is bounded by $\int dt~a^3 \rho_{\phi}$ from above. We can then consider the quantity
\begin{equation}
I_1=\frac{\int_{t_i}^T dt~a^3 \rho_{\phi}}{\int_{t_i}^T dt~a^3}.
\end{equation}
Because $a(t)$ increases with time, we only need to check the late-time behavior of $\rho_{\phi}$. In a typical quintessence/inflation scenario, we have $V_{\min}=0$ and $\rho_{\phi}\propto a^{-n}$ with $n(t)=\frac{3}{1+w_{\phi}(t)}>0$ throughout the evolution. Thus, in the $T\rightarrow \infty$ limit, $I_1\rightarrow 0$. This establishes $\bar{S}=0$, implying that any such scenario remains untouched in the volume-normalized framework. One can also freely include ordinary matter ($\rho \propto a^{-3}$) or radiation ($\rho \propto a^{-4}$) in the scenario, because they do not contribute to $\bar{S}$, by the same argument.

Note that a constant shift of $V(\phi)$ does not change the Friedmann equations. A straightforward generalization of the analysis above shows that it is only the ``non-constant" part of $\rho_{\phi}$, $V-V_{\min}$, that drives cosmic expansion. While this follows intuitively from the cosmological-constant-freedom of the volume-normalized action, it can be seen nontrivially in (\ref{sbar1}): Because $V(\phi(T))-V_{\min}=0$ by definition and the integrals are dominated by late-time contributions, the first term of (\ref{sbar1}) vanishes for any sensible $V(\phi)$ and $\phi$ profiles, and the sole effect of $\bar{S}$ is to neutralize a cosmological constant.

As an explicit example, consider the ultralight pseudo-Nambu-Goldstone boson (PNGB) model for dark energy \cite{pngb}, in which the PNGB scalar field has the effective potential
\begin{equation}\label{p1}
V(\phi)=M^4\left[1- \cos\left( \frac{\phi}{f}\right)  \right].
\end{equation}
Here, $M \sim $ light neutrino mass $\sim 10^{-2}-10^{-3}~\text{eV}$, and $f \sim$ global symmetry breaking scale $\sim 10^{16}-10^{19}~\text{GeV}$.
The structure of this potential is protected by the fermionic chiral symmetries, so the smallness of $m_{\phi} \sim \frac{M^2}{f} \sim 10^{-29}-10^{-34}~\text{eV}$ is natural from the particle physics point of view. At early times, $\rho_{\phi} \simeq -p_{\phi} \sim M^4$ mimics a CC; at late times, $\rho_{\phi} \sim a^{-3}$ looks like ordinary matter \cite{pngb}. This clearly guarantees $\bar{S}=0$, and thus represents a viable and theoretically favorable dark energy scenario in the volume normalized framework. The same can be said about the natural inflation model, where the QCD axion PNGB has the same potential (\ref{p1}), except that it probes the GUT scale, $M \sim 10^{16}$ GeV and $f \gsim 10^{19}$ GeV \cite{ni}. 

On the flip side, let us also briefly look at a minimally coupled phantom scalar \cite{psca}, in which the scalar field has the wrong (flipped!) sign kinetic term. For a homogeneous phantom scalar on flat FRW background, $\rho_{\phi}=-\frac{1}{2}\dot{\phi}^2+V(\phi)$, $p_{\phi}=-\frac{1}{2}\dot{\phi}^2-V(\phi)$, $w_{\phi}<-1$, and at late times, $\rho_{\phi}\propto a^n$ with $n>0$. Following exactly the aforementioned argument yields
\begin{equation}
-\bar{S}=\frac{\int dt~a^3 V(\phi(t))}{\int dt~a^3}>\frac{\int dt~a^3 \rho_{\phi}}{\int dt~a^3}=\infty,
\end{equation}
if one uses a $\phi$ profile that leads to cosmic expansion. This is nonsensical in view of the first Friedmann equation (\ref{f1}), so phantom energy is ruled out in the volume-normalized framework.

\section{Nonminimally Coupled Scalar} \label{nmcs}
A scalar field nonminimally coupled to gravity can be described in the volume-normalized framework by
\begin{equation}\label{nm1}
\bar{S}=\frac{S}{V_g}=\frac{\int d^4x~\sqrt{g}\left[\frac{M_0^2+f(\phi)}{2}R+\mathcal{L}_{\rm m}(g^{\mu \nu},\phi,\psi_I) \right]}{\int d^4x~\sqrt{g}},
\end{equation}
where $M_0$ is some mass parameter, and $\mathcal{L}_{\rm m}$ is the matter Lagrangian that includes $\phi$ and the (other) Standard Model (SM) fields $\psi_I$. In principle, $M_0$ could be set to zero, under which the vacuum expectation value (VEV) of $f(\phi)$ then induces gravity. Here we keep the general identification, $M_{\rm P}^2=M_0^2+\langle f(\phi) \rangle_{\rm VEV}$. In this convention, it is the metric $g_{\mu \nu}$ and its inverse that all the fields in $\mathcal{L}_{\rm m}$ are coupled to; for example,
\begin{equation}
\mathcal{L}_{\rm m}(g^{\mu \nu},\phi,\psi) \supset -\frac{1}{2}g^{\mu \nu} \partial_{\mu}\phi \partial_{\nu} \phi -V(\phi).
\end{equation}
We emphasize this point because it makes the form (\ref{nm1}) unique if we are to retain the solution of the cosmological constant problem. In general, because of the conformal factor in the numerator, one could introduce scalar field dependence in the denominator in (\ref{nm1}) when defining a scalar-tensor theory in the volume normalized framework of (\ref{ac2}). However, the matter quantum loops will generate a cosmological constant term in the numerator, $\int d^4x\sqrt{g} \Lambda_{\rm m}$, with $\Lambda_{\rm m}$ being a veritable constant independent of the fields. In order for this quantum correction to CC to drop out of the physical picture via (\ref{ac2}),
one really needs to keep the volume factor in the denominator 
independent of the scalar field; this selects the form (\ref{nm1}). We also call the fields written in (\ref{nm1}) in the ``Jordan frame," to relate to the well known nomenclature.    

Then the Einstein field equations become
\begin{equation}
\left[ M_0^2+f(\phi) \right]G_{\mu \nu}=T_{\mu \nu}-\bar{S}g_{\mu \nu}+\left( \nabla_{\mu}\nabla_{\nu}-g_{\mu \nu}\square \right)f(\phi),
\end{equation}
where $T_{\mu \nu}$ is defined as usual,
\begin{equation}
T_{\mu \nu}=\frac{-2}{\sqrt{g}}\frac{\partial(\sqrt{g}\mathcal{L}_{\rm m}(g^{\mu \nu},\phi,\psi))}{\partial g^{\mu \nu}}.
\end{equation}
Using only $\mathcal{L}_{\rm m}(g^{\mu \nu},\phi,\psi) \rightarrow -\frac{1}{2}g^{\mu \nu} \partial_{\mu}\phi \partial_{\nu} \phi -V(\phi)$, we continue to write the Friedmann equations (with the replacement $M_{\rm P}\rightarrow M_0$) in the form of (\ref{f1}) and (\ref{f2}), but now with 
\begin{gather}
\rho_{\phi} = \frac{1}{2}\dot{\phi}^2+V(\phi)-3Hf'(\phi)\dot{\phi},\\
p_{\phi}=\frac{1}{2}\dot{\phi}^2-V(\phi)+ 2Hf'(\phi) \dot{\phi}+f'(\phi)\ddot{\phi}+f''(\phi)\dot{\phi}^2,
\end{gather}
and the $\phi$ equation becomes
\begin{equation}\label{phi2}
\ddot{\phi}+3H\dot{\phi}+V'(\phi)-3(\dot{H}+2H^2)f'(\phi)=0.
\end{equation}
Plugging the Friedmann equations back into (\ref{nm1}),
\begin{align}
\bar{S}&=\frac{1}{\int dt~a^3}\int dt~a^3 \left\{  -V(\phi)+ \frac{3}{2} \left[ 3Hf'(\phi) \dot{\phi}+f'(\phi)\ddot{\phi}+f''(\phi)\dot{\phi}^2 \right] \right\} \label{line1}\\
&=-\frac{1}{\int dt~a^3}\int dt~a^3 \left\{ \rho_{\phi} -\frac{1}{2}\dot{\phi}^2- \frac{3}{2} \left[ Hf'(\phi) \dot{\phi}+f'(\phi)\ddot{\phi}+f''(\phi)\dot{\phi}^2 \right] \right\}.  \label{line2}
\end{align}
Compared to (\ref{sbar1}), the new terms in the integrand of (\ref{line1}) are all proportional to the time derivatives of $\phi$, which vanish at the end of the cosmic evolution. Then we have $\bar{S}=-V_{\min}$ by the same argument in the last section for reasonable potentials and $\phi$ profiles. Also, note that by (\ref{line2}), for $\rho_{\phi}$ proportional to the negative powers of $a(t)$, which necessarily means $V_{\min}=0$, we have $\bar{S}=0$ identically. Therefore, the classical pictures of cosmic expansion in the nonminimally coupled models are exactly the same in the frameworks of $S$ and $\bar{S}$.

This result applies generically to a very wide variety of nonminimally coupled dark energy \cite{nmcintro,nmcintro1,nmcintro2}\footnote{See also, \cite{nmc21,nmc22,nmc223,nmc23,nmc24}.} and inflation models \cite{infnmc01,infnmc02}.\footnote{See also, \cite{infnmc03,infnmc21,infnmc22,infnmc23,infnmc24}.} Notably, the Higgs inflation model uses the radial mode of the Higgs doublet as the nonminimally coupled scalar, needing no extra field beyond the Standard Model \cite{higgs1,higgs2}. There, the equations of motion receive extra contributions from the coupling between Higgs and other SM fields, but these local effects clearly do not affect the argument above.

So far we have worked in the Jordan frame. In the nonminimally coupled scalar setup, it is also popular to transform to the Einstein frame, in which we will demonstrate the result above with the choice \cite{higgs1,higgs2}
\begin{equation}
f(\phi)= \xi \phi^2, \quad V(\phi)=\frac{\lambda}{4}\left( \phi^2-v^2 \right)^2.
\end{equation}
For concreteness, we take $\phi$ to be the radial Higgs mode, and then $M_0 \approx M_{\rm P}$ with very high accuracy. Following \cite{higgs1,higgs2}, the field redefinitions read
\begin{equation}
g_{\mu \nu}=\Omega^{-2}\hat{g}_{\mu \nu},\quad \Omega^2=1+\frac{\xi \phi^2}{M_{\rm P}^2},\quad \quad \frac{d\chi}{d\phi}=\sqrt{\frac{\Omega^2+\frac{6\xi^2\phi^2}{M_{\rm P}^2}}{\Omega^4}},
\end{equation}
upon which (\ref{nm1}) with $\mathcal{L}_{\rm m}(g^{\mu \nu},\phi,\psi) \rightarrow -\frac{1}{2}g^{\mu \nu} \partial_{\mu}\phi \partial_{\nu} \phi -V(\phi)$ becomes
\begin{equation}\label{nmcs2}
\bar{S}=\frac{\int d^4x~\sqrt{\hat{g}}\left[\frac{M_{\text P}^2}{2}\hat{R}-\frac{1}{2}\hat{g}^{\mu \nu}\partial_{\mu}\chi\partial_{\nu}\chi -U(\chi) \right]}{\int d^4x~\Omega^{-4}(\chi)\sqrt{\hat{g}}},
\end{equation}
where $\hat{R}$ is calculated from $\hat{g}_{\mu \nu}$, and
\begin{equation}
U(\chi)=\Omega^{-4}(\chi)\frac{\lambda}{4} \left( \phi^2(\chi)-v^2 \right)^2.
\end{equation}
On rolling solutions that lead to cosmic expansion, the field values are large at early times and small at the end of the evolution, so $\Omega^{-4}$ gradually increases to 1. Then the situation is essentially reduced to that of a minimally coupled scalar $\chi$ with $U_{\min}=0$, because the integrals in (\ref{nmcs2}) are clearly dominated by late-time contributions. Repeating the analysis in the last section step by step then yields $\bar{S}=0$, as expected.

\section{Chaplygin Gas}\label{cg}
Chaplygin gas is an exotic cosmic fluid that has the equation of state $p=-\frac{A}{\rho}$. In a flat FRW universe, energy conservation implies that
\begin{equation}\label{chap1}
\rho = \sqrt{A+\frac{B}{a^6}},
\end{equation}
which represents a unified dark energy/matter fluid that transitions from a CC to a ``stiff matter" with $\rho \sim a^{-6}$ \cite{chap}. Microscopically, it has been shown that the Chaplygin gas equation of state can be induced by a 3-brane living in a 5D Minkowski space on the 4D braneworld transverse to its lightcone \cite{microchap-1,microchap0}.\footnote{See also, \cite{microchap1,microchap2,microchap3}.} It remains to be checked if this picture can be made compatible with (\ref{pr1}). Anyway, we present the Chaplygin gas scenario as a negative example.

The profile (\ref{chap1}) can be realized by a scalar field minimally coupled to gravity with the potential
\begin{equation}
V(\phi)=\frac{\sqrt{A}}{2}\left[ \text{cosh}\left( \frac{\sqrt{3}\phi}{M_{\rm P}}\right) +\frac{1}{ \text{cosh}\left( \frac{\sqrt{3}\phi}{M_{\rm P}}\right) } \right].
\end{equation}
However, in the volume normalized framework, the analysis in section \ref{mcs} shows that $V_{\min}=\sqrt{A}$ does not contribute to the Friedmann equations. The effective potential in this case is
\begin{equation}
V_{\rm eff}=V(\phi)-V_{\min}=\frac{\sqrt{A}}{2}\left[ \text{cosh}\left( \frac{\sqrt{3}\phi}{M_{\rm P}}\right) +\frac{1}{ \text{cosh}\left( \frac{\sqrt{3}\phi}{M_{\rm P}}\right) } -2 \right],
\end{equation}
so the Chaplygin gas solution is no longer valid. In fact,
\begin{equation}
V_{\rm eff}\propto \frac{ \left[ \text{cosh}\left( \frac{\sqrt{3}\phi}{M_{\rm P}}\right)-1\right]^2}{\text{cosh}\left( \frac{\sqrt{3}\phi}{M_{\rm P}}\right)},
\end{equation}
which describes just another quintessence scenario interpolating between known models, $V(\phi) \propto \left[ \text{cosh}\left( \frac{\sqrt{3}\phi}{M_{\rm P}}\right)-1\right]^q$ with $q=1,2$ \cite{quint1}. The same argument here also applies to generalized Chaplygin gas models \cite{gchap}, simply because the extra terms in the Friedmann equations nullify any CC-like component.

\section{Fluctuations and Quantum Effects in Inflation}\label{conc}
We first make some general observations about the volume normalized action, (\ref{ac2}). Its variational procedure yields
\begin{equation}
\delta \bar{S}=\delta\left( \frac{S}{V_g}\right)=\frac{1}{V_g}\left(\delta S-\frac{S}{V_g}\delta
V_g \right),
\end{equation}
so all nongravitational local physics encoded in $S$ are left unchanged, while the Einstein field equations are modified by the extra term $\propto \bar{S}$. The latter is then the key to understand the differences between the theories described by $S$ and $\bar{S}$. In both dark energy and inflationary scenarios,  the scale factor $a(t)$ increases with time and becomes very large at the end of the evolution, so $S$ is dominated by the late-time behavior of its integrand, which implies that, $\bar{S}=-V_{\min}$. Without surprise, this is just a roundabout way to see that the volume normalized action automatically gets rid of any CC-like contribution to the cosmic equation of state. 

We then move on to field fluctuations. For the scalar fluctuations $\delta \phi(\vec{x},t)$, both classical and quantum, they are merely local effects and do not contribute to $\bar{S}$. The same can be said about the metric fluctuations $\delta  g_{\mu \nu}(\vec{x},t)$. Finite temperature has also no effect on $\bar{S}$, which depends only on the late-time picture of the Universe that approaches a zero temperature state. Thus, the dynamics of field fluctuations do not differ from that obtained via $S$, except that $\delta  g_{\mu \nu}(\vec{x},t)$ are effectively classical, owing to the vanishing Planck constant for gravity loops, implied by the full action (\ref{pr1}). We will see below the nontrivial effects of this fact in the inflationary cosmology.

Because the inflaton is still quantized with $\hbar$, the inflationary predictions for primordial scalar modes do not change in the volume-normalized framework. This can be seen in two ways. First, even though the scalar perturbation is a mixture of the inflaton and the metric fluctuation, it is  possible to choose a gauge in which the scalar modes are due entirely to the 
inflaton fluctuations. Second, one needs not choose a gauge, but rather organize the scalar perturbations into gauge invariant quantities, which are to be diagonalized and canonically normalized before quantization. For example, during the de Sitter stage of a minimally coupled model, the quadratic action for scalar perturbations can be reduced to \cite{infreview1,infreview3,infreview4,infreview5}
\begin{equation}\label{muk}
S_2=\frac{1}{2}\int d^3xd\eta\left[ (u')^2-(\partial_iu)^2+\frac{z''}{z}u^2 \right],
\end{equation} 
where $\eta$ is the conformal time and primes are derivatives with respect to it. The gauge invariant quantities are
\begin{equation}
u=z\mathcal{R},\quad \mathcal{R}=\left( \Psi+\frac{H}{\dot{\phi}}\delta \phi \right), \quad z^2=\frac{a^2\dot{\phi}^2}{H^2},
\end{equation}
where $\mathcal{R}$ is known as the comoving curvature perturbation, and $\Psi$ is defined in the way that the scalar and tensor metric fluctuations are expressed as
\begin{equation}\label{fluc}
ds^2=-(1+2\Phi)dt^2+2a\partial_iBdx^idt+a^2\left[ (1-2\Psi)\delta_{ij}+ 2\partial_i \partial_j E+h_{ij} \right]dx^idx^j.
\end{equation} 
One can then go to a gauge with $\Psi=0$, so that $u$ is due to $\delta \phi$ only. One can also directly quantize $u$ from (\ref{muk}), without fixing $\Psi$ or $\delta \phi$. In the latter case, the quantized scalar modes are completely equivalent regardless of whether the scalar metric fluctuations are fundamentally classical or quantum, thus leading to the same predicted power spectra. The subtleties of diagonalizing a quantum field and a classical field are addressed in Appendix \ref{2scalar}.

Curiously, the tensor metric fluctuations $h_{ij}$ in (\ref{fluc}), which satisfy the transverse traceless 
condition $\partial^ih_{ij}=h^i_i=0$,  and evolve separately, are not quantized in the volume-normalized 
framework, thus giving vanishing quantum correlation functions and power spectra. This implies that no primordial 
tensor modes arise during inflation -- that is, the theory predicts a zero tensor-to-scalar ratio, $r=0$, irrespective of the inflationary model. Thus,  the framework (\ref{pr1}) with $\hbar V_g/V_F\to 0$ could be falsified  
by observational data that would deny this null prediction, for example, by the detection of the primordial 
gravitational waves.

\chapter{Biverse Volume-Normalized Massive Gravity} \label{ch4}
This chapter is based on the work: Gabadadze, G., and Yu, S. (2016). Metamorphosis of the Cosmological Constant and 5D Origin of the Fiducial Metric. {\it Phys. Rev. D}, 94:104059. [arXiv:1510.07943 [hep-th]].\footnote{Copyright (2016) by the American Physical Society, \url{https://doi.org/10.1103/PhysRevD.94.104059}.}
\\

In this chapter, we choose massive gravity to generate dark energy in the volume-normalized framework, assuming that our $g$-universe lives on the conformal boundary of an AdS$_5$ bulk. We describe the resulting classical and quantum aspects in detail. In particular, we will see that the quantization scheme described in Chapter \ref{ch2} removes the problem of otherwise 
unmanageable counterterms of massive gravity at its strongly coupling scale.

\section{A New Embedding}

The construction of the biverse volume-normalized action invokes  two  assumptions: 
(I) the CC in the $F$-universe was assumed to be  large and negative, without 
tunings of any kind;  (II) dark energy in the $g$-universe  was postulated to be 
due to massive gravity.  Both assumptions seem to be of  
a provisional character, as one could come up with other  scenarios 
giving $V_g/V_F \ll 1$, with dark energy produced by a different technically 
natural mechanism.

Having said this,  however,  we note that assumption (I) may have a strong justification:  
the $F$-universe with its negative CC  could  be supersymmetric, and  broken supersymmetry  
in the $g$-universe needs not be communicated to the $F$-universe,  
since $F$ and $g$  are only globally connected \cite {Gabadadze:2014rwa}.

How about assumption (II)? It is not necessary, but could there 
be advantages in using massive gravity, as opposed to other 
forms of dark energy (CC excluded)? Are there specific features of massive gravity 
that naturally fit into the above proposal for solving the big CC problem?  If the answers to these questions were positive, then the proposed mechanism   for the CC problem  would be connected 
to the mechanism for dark energy.  

In this chapter, we claim partial success in this quest. In particular, we will show  
how the fiducial metric -- a necessary  ingredient  of the diffeomorphism 
invariant  formulation   of massive gravity --   can be related 
to the metric of the $F$-universe, introduced to solve the big CC problem. 
Our construction unveils a higher dimensional origin of the fiducial metric 
in massive gravity, and gives more appeal  to the scenario in which dark energy is due to 
massive gravity  and its extensions.\footnote{See, e.g., \cite {Mukohyama,Tanaka} and references therein.} 

As a by-product, we  will show that embedding of massive gravity 
into the action $S$ in (\ref {acmain})  helps deal with  a well-known  
difficulty of this theory  related to its  quantum strong coupling behavior 
at the scale  $\Lambda_3 =(M_{\rm Pl} m^2)^{1/3}$: the quantum loop corrections to 
the massive gravity action are now controlled by 
$\hbar q \to 0$.  Therefore,  the  loop diagrams do not generate any counterterms, 
beyond the terms already present in the tree-level  action.    This 
does not eliminate  the  scale $\Lambda_3$,  and the theory is still 
strongly coupled at that scale.   However, the effective action receives no 
loop-generated counterterms.\footnote{The tree  
amplitudes do determine the respective loop amplitudes via the optical theorem, which 
remains valid order by order in $\hbar q$. See discussions in Section \ref{secstrong}.}  
Thus,  the full quantum effective action has as few terms as the 
tree-level action, and calculations done in this theory -- albeit strongly coupled -- are  exact, unlike in a generic non-renormalizable  theory  that would  generate an infinite number of  new counterterms.

The  main idea  here is to regard   the action of the ``other" universe as the EH 
action of  5D gravity,  described by the metric $F$ on  AdS$_5$. We then endow a hypersurface at the AdS$_5$ boundary  with the dynamical metric $g$,  and  
couple $g$ to the pullback of the bulk metric,  $\gamma$,  in such a way  that the latter acts as the fiducial metric of 4D massive gravity described  by $g$.

The pullback $\gamma $ can be expressed via  the  four St\"uckelberg  
scalars \cite{AGS} of massive gravity $\varphi^a(x)$,  ${ \gamma}_{\mu\nu} = \partial_\mu \varphi^a 
\partial_\nu \varphi^a  \eta_{ab} ~(a,b,... =0,1,2,3)$. These scalars  
then end up parametrizing the 4D hypersurface at the  boundary of the AdS$_5$ bulk.
The detailed construction is worked out in  Section \ref{secm1}, where a simple local 
model without the $V_F/V_g$ factor is considered.  Besides its illustrative purpose, this model also 
delineates a method of introducing gravity, albeit massive, on the boundary of AdS$_5$. 
The full model (\ref{acmain}) is examined in Section \ref{secm2}, where we show 
how the fiducial metric of massive gravity emerges from the pullback 
of the bulk metric in AdS$_5$. The discussion of the 
strong coupling problem of massive gravity will be given in Section \ref{secstrong}.

\section{Massive Gravity On Top of the Boundary of AdS$_5$}\label{secm1}

Our bottom-up construction  begins with postulating a dynamical 4D metric,  $g_{\mu\nu}$, 
on top of the boundary of AdS$_5$. The metric  $g$  is not related to  a pullback of the bulk gravity.  
Such a  boundary has more structure than  
the conventional conformal boundary of AdS$_5$, and may be thought as 
some geometrical hypersurface, that is 
endowed with  $g$ and  is placed right at the boundary  of AdS$_5$.

To set the conventions, $x^{\mu}$ (with $\mu, \nu,...=0,1,2,3$)  denote the  coordinates  of 
4D spacetime of the hypersurface  that is endowed with the  dynamical metric $g_{\mu\nu}(x)$.
The massive gravity  action  defined via $g$  on the world-volume of the boundary 
hypersurface reads as follows, 
\begin{equation}\label{mg1}
S={M^2_{\rm P}} \int d^4x \sqrt{g}\left[ R(g)-2\Lambda+2m^2\mathcal{U} \left(\sqrt{g^{-1}\gamma } \right) \right],
\end{equation}
where  the  diffeomorphism invariant potential  $\mathcal{U}$  was built in \cite{deRham:2010ik,deRham:2010kj}, 
as a function of the metric $g$,  and the fiducial Minkowski metric ${ \gamma}_{\mu\nu} = \partial_\mu \varphi^a 
\partial_\nu \varphi^a  \eta_{ab} ~(a,b,... =0,1,2,3)$, in an arbitrary 
coordinate system;    this potential  can be written in  the following form:
\begin{equation}\label{mGRUK0}
\mathcal{U}(\mathcal{K})=\det_2(\mathcal{K})+\alpha_3\det_3(\mathcal{K})+\alpha_4\det_4(\mathcal{K}),
\end{equation}  
where  the matrix $\mathcal{K}=1-A$, with $A$ being  defined  as one of the roots of,  
$A^{\mu}_{\alpha}A^{\alpha}_{\nu}=g^{\mu \alpha}\gamma_{\alpha \nu}$, so that 
$\mathcal{K}=1-\sqrt{g^{-1}\gamma}$.

The fiducial metric $\gamma$ is not dynamical,  and its origin is unknown in massive gravity.  
This issue is clarified in the present approach  by relating the fiducial metric to 
the pullback of  the  bulk metric. To see this relation
we define the bulk coordinates in AdS$_5$ as  $Y^A$'s with $A,B,...=0,1,2,3,5$.  
The 5D and 4D coordinates are related as follows, 
\begin{equation}
Y^{A}=\{\varphi^a(x),z \}.
\end{equation}
In this coordinate system,   the 4D boundary hypersurface is located at $z=0$, 
while the fields  $\varphi^a(x)$ (with $a,b,...=0,1,2,3$) give 
different  parametrizations  of the hypersurface.  
As to the bulk action,  it is the 5D EH action with the negative CC and 
with the Gibbons-Hawking  terms included (see, e.g., \cite{Liu:1998bu}),
\begin{equation}\label{s5}
S_F=\int_{M} d^5Y~\sqrt{F}\left(R(F)+\frac{12}{l^2} \right)+2\int_{\partial M}d^4Y~\sqrt{{\tilde \gamma}}K-\frac{6}{l}\int_{\partial M}d^4Y~\sqrt{{\tilde \gamma}}.
\end{equation}
Here, the 5D Planck mass $M_5$ is set to unity, $\tilde \gamma$ is the induced metric on the boundary, and $K$ is the trace of the boundary extrinsic curvature.  Our ultimate goal is to study the functional 
\be \label{SSF}
{\cal A}_5 = {V_F\over V_g} \, S \,+\,S_F,
\ee
but in this section, as a warm-up exercise,  we study  (\ref {SSF}) without 
the $V_F/V_g$ factor in front of $S$. Thus, the total action of  the  theory  considered in 
this section is 
\begin{equation}\label{acm1}
S_{\rm tot}=S+S_F.
\end{equation}
The action (\ref {s5})  contains a negative cosmological constant, 
$-\frac{6}{l^2}$, and thus generates an $\text{AdS}_5$ solution to the respective Einstein equations,
\begin{align}\label{m1}
ds^2&=F^{(0)}_{AB}dY^AdY^B \nonumber \\ &=\frac{l^2}{z^2}\left(\eta_{ab}d\varphi^{a}d\varphi^{b}+dz^2 \right) \\ &=\frac{l^2}{z^2}\left(\eta_{ab}\partial_{\mu}\varphi^{a}\partial_{\nu}\varphi^{b}dx^{\mu}dx^{\nu}+dz^2 \right). \nonumber
\end{align}
The interval in the last line  is written in terms of the boundary massive gravity coordinates. 

We are now in a position to relate the induced  metric on the boundary, $\tilde{\gamma}$, to  
the fiducial metric of massive gravity, $\gamma$.   Our choice for this relation reads as
\begin{equation}\label{bm1}
\gamma _{\mu \nu}(x)=\lim_{z\rightarrow 0} {z^2\over l^2}{\tilde \gamma}_{ab}(z,\varphi(x))\frac{\partial \varphi^a}{\partial x^{\mu}}\frac{\partial \varphi^b}{\partial x^{\nu}}.
\end{equation}
This  identifies the St\"ukelberg fields used in the  massive gravity potential as the coordinates of the 
boundary in the AdS$_5$ bulk. 
The variational principle  for the total action is defined as follows:  $S_F$ is varied w.r.t. $F$. 
The boundary action, $S$,  on the other hand,  is  varied w.r.t. $g$, but no variation is taken w.r.t. $\gamma$.
The induced bulk  metric on the boundary, $\tilde \gamma$,  and the fiducial metric, $\gamma$,
are then related to each other via the boundary condition  (\ref {bm1}).  
Under this premise, the linearized theory with the Dirichlet boundary data is studied 
next.\footnote{We use $S_F$ as the bulk action even though it contains the  Gibbons-Hawking 
boundary terms, the sole role of which is to guarantee  the right bulk  
Einstein equations.   In contrast,  the action $S$, referred here as the boundary action, 
gives rise to the dynamical equations of motion on the boundary.  We hope this nomenclature 
will not cause any serious confusion.}

The derivation of the on-shell boundary effective action  from  the bulk action, $S_F$, 
is a well-known calculation in standard AdS/CFT. Here we give an outline of this procedure 
adapted to our action (\ref{acm1}). For details, we refer the readers to Appendix \ref{append1}.

In the following, we will  work at the regularized boundary $z=\epsilon$,  and  the limit $\epsilon\rightarrow 0$  will be 
taken only at the end of the calculations. In the linearized theory, the full bulk metric is $F_{AB}= F_{AB}^{(0)}+h_{AB}$, where $h_{AB}$ denote the components of 5D metric fluctuations. We work in the gauge $h_{0B}=0$ and introduce $h_{ab}(z,\varphi)=z^2h_{AB}\delta^A_a\delta^B_b/l^2$. Then we can  write
\begin{equation}
ds^2=\frac{l^2}{z^2}\left[(\eta_{ab}+h_{ab})d\varphi^{a}d\varphi^{b}+dz^2 \right].
\end{equation}
By (\ref{bm1}), the fiducial metric on the boundary receives corrections from the 5D fluctuations,
\begin{equation}
\gamma_{\mu \nu}(x)=\eta_{ab}\partial_{\mu}\varphi^a\partial_{\nu}\varphi^b+\delta\!\gamma_{\mu \nu}(x),
\end{equation}
 where
\begin{equation}
\delta\!\gamma_{\mu \nu}(x)=\lim_{\epsilon\rightarrow 0} h_{ab}\left(\epsilon,\varphi(x)\right)\partial_{\mu}\varphi^a \partial_{\nu}\varphi^b.
\end{equation}
Thus, $\delta\!\gamma_{\mu \nu}$ represent the induced fluctuations. We also define the related metric $\gamma_{ab}(\varphi(x))=\eta_{ab}+h_{ab}(\epsilon,\varphi(x))|_{{\epsilon\rightarrow 0}}$, so that $\gamma_{\mu \nu}(x)=\gamma_{ab}(\varphi(x))\partial_{\mu}\varphi^a \partial_{\nu}\varphi^b$. The resulting  effective boundary 
action can be written as a sum of various terms
\begin{equation}
S[h_{ab}]=S_{\partial^2}+S_{\partial^4}+S_{\text{nonlocal}},
\end{equation}
where the respective parts of the total action are defined as follows:
\begin{equation}\label{dd2}
S_{\partial^2}\sim ~\frac{1}{\epsilon^2}\int d^4\varrho ~h\partial^2h,\quad S_{\partial^4}\sim ~\ln \epsilon \int d^4\varrho~h\partial^4h,
\end{equation}
\begin{equation}
 S_{\text{nonlocal}}\sim \int d^4\varrho d^4\varsigma~\frac{h(\epsilon,\varrho)h(\epsilon,\varsigma)}{|\varrho-\varsigma|^8}.
\end{equation}
The exact formulae and tensor structures can be found in (\ref{d2}) -- (\ref{nl1}). In particular, the nonlocal terms give the two-point function for the boundary CFT stress-tensor,

\begin{equation}
S_{\text{nonlocal}}=-\frac{1}{2}\int d^4\varrho d^4\varsigma ~h_{ab}(\epsilon,\varrho)h_{cd}(\epsilon,\varsigma)\Expect{\T^{ab}(\varrho)\T^{cd}(\varsigma)},
\end{equation}
\begin{align}
\Expect{\T^{ab}(\varrho)\T^{cd}(\varsigma)}\equiv~&\frac{20}{\pi^2}\frac{1}{|\varrho-\varsigma|^8} \nonumber \\
&\times \left[\frac{1}{2} J^{ad}(\varrho-\varsigma)J^{bc}(\varrho-\varsigma)+\frac{1}{2}J^{ac}(\varrho-\varsigma)J^{bd}(\varrho-\varsigma)-\frac{1}{4}\eta^{ab}\eta^{cd} \right], \nonumber
\\ J^{ab}(\varrho)=&\eta^{ab}-2\frac{\varrho^a \varrho^b}{|\varrho|^2}.
\end{align}
As a consistency check, the $S_{\partial^2}$ terms comprise the standard kinetic terms for linearized Einstein gravity,
\begin{equation}\label{kin1}
S_{\text{kin}}[h_{ab}]=\frac{1}{2\epsilon^2}\int d^4\varrho~\left[\frac{1}{4}h_{ab}\square h^{ab}-\frac{1}{2}h_{ab}\partial^a\partial_ch^{bc}-\frac{1}{4}h\square h+\frac{1}{2}h_{ab}\partial^a\partial^b h \right].
\end{equation}

The  local terms in (\ref{dd2}) with divergent powers of $\epsilon$ are cancelled by appropriate counterterms introduced as part of the renormalization procedure. This procedure deprives the fields $h_{ab}(\epsilon,\varphi)$ of any dynamics. As a result, one 
regards  them as the boundary Dirichlet  data supplied by specified boundary conditions. 
We highlight this fact by calling $h^0_{ab}(\varphi(x))=h_{ab}(\epsilon,\varphi(x))|_{\epsilon\rightarrow 0}$, so that $\gamma_{ab}(\varphi(x))=\eta_{ab}+h^0_{ab}(\varphi(x))$. Then the total effective boundary action is
\begin{align}\label{mGRunmod}
S^{\rm eff}_{\text{tot}}[g,f]=&{M_{\rm Pl}^2}\int d^4x\sqrt{g}\left[ R(g)-2\Lambda+2m^2\mathcal{U}\left( \sqrt{g^{-1}\gamma_{ab}(\varphi(x))\partial_{\mu}\varphi^a \partial_{\nu}\varphi^b} \right) \right] \nonumber \\
&-\frac{1}{2}\int d^4\varrho d^4\varsigma~h^0_{ab}(\varrho)h^0_{cd}(\varsigma)\Expect{\T^{ab}(\varrho)\T^{cd}(\varsigma)}.
\end{align} 
 Likewise, the corrections to the fiducial metric can be identified with the Dirichlet boundary data up to a 
 diffeomorphism,
\begin{equation}
\delta\!\gamma_{\mu \nu}(x) =h^0_{ab}(\varphi(x))\partial_{\mu}\varphi^a \partial_{\nu}\varphi^b.
\end{equation}
Thus, both the CFT and the massive gravity sectors are sourced by the same boundary fields. If the 
standard AdS/CFT prescription is applied to the above action, (\ref {mGRunmod}), the CFT correlation 
functions will be modified by terms that are proportional to powers of the graviton mass. Thus,  for a small graviton mass, 
one would arrive at a softly broken CFT, but this is not our concern for the time being.  

\section{Removing CC and Introducing  Dark Energy}\label{secm2}

We now turn to the action functional that in the present context 
can remove an arbitrary 4D CC from the  dynamical 4D boundary of 5D AdS gravity. 
In the spirit of (\ref{acmain}), we look at the action (\ref{SSF}),
\begin{equation}\label{modaction}
{\cal A}_5=\frac{V_F}{V_g}S+S_F.
\end{equation}  
Here, $V_F=\int d^5Y \sqrt{F}$ and $V_g=\int d^4x\sqrt{g}$. All the conventions here are the same as in 
the previous sections. To avoid ambiguity in the equations of motion due to ratios of infinite volumes, we regularize all integrals in (\ref{modaction}),
\begin{align}
\int d^4x &= \lim_{\bar{x}^0,...,\bar{x}^3\rightarrow \infty} \left(\prod_{\mu=0}^3 \int_{-\bar{x}^{\mu}}^{\bar{x}^{\mu}} \right) d^4x \equiv \lim_{\bar{x}\rightarrow \infty}\int_{\text{reg}}d^4x ,\nonumber \\
\int d^5Y &=\int dz~d^4\varphi= \lim_{\bar{\varphi}^0,...,\bar{\varphi}^3\rightarrow \infty} \lim_{\epsilon \rightarrow 0}  \int_{\epsilon}^{\infty}dz \left( \prod_{A=0}^3 \int_{-\bar{\varphi}^A}^{\bar{\varphi}^A} \right) d^4\varphi~ \equiv \lim_{\bar{Y}\rightarrow \infty} \lim_{\epsilon \rightarrow 0} \int_{\text{reg}}d^5Y,
\end{align}
so (\ref{modaction}) is actually
\begin{equation}
{\cal A}_{5}=\lim_{\bar{x},\bar{Y}\rightarrow \infty } \lim_{\epsilon \rightarrow 0} \left[ \frac{V_F}{V_g}S+S_F\right]_{\text{reg}}.
\end{equation}
The order of  the limits is carefully arranged: $\epsilon\rightarrow 0$ is taken before $\bar{x},\bar{Y}\rightarrow \infty$. This regularization first restricts the theory to a 5D ``box," then extends the $z$ direction to include the conformal boundary, before taking  the remaining  limits. For conciseness, we 
will always treat the volumes as regularized, but will not write the  regularizations explicitly.

From the discussion in Chapter \ref{ch2}, it is clear that in the action (\ref {modaction}), the 4D CC does not curve the 4D spacetime, 
but instead gets absorbed into the 5D CC. If the magnitude of the 5D CC is greater 
than that of the 4D CC, then this leads to  an obvious modification of the curvature of the 
AdS bulk, while all the conclusions of the previous section remain valid.  An order-of-magnitude  hierarchy between
the scales of the 5D and 4D CC's is straightforward to arrange, since  5D theory can be 
exactly supersymmetric, while the broken SUSY in 4D could guarantee its scale to be lower. Moreover, since $V_F$ now is the five-volume of 
the 5D AdS spacetime, and $V_g$ is going to be  the  four-volume 
of the 4D dS spacetime, the ratio $V_g/V_F \to 0$ in Planck units. This guarantees 
that the outlined classical solution is stable w.r.t. quantum loops.

What remains to be seen is how the fiducial metric 
arises in  the boundary massive gravity, and how the theory gives rise to dark energy. 
To this end, we just follow, step by step, the procedure of ``integrating out" the AdS 
bulk described in Chapter \ref{ch2}. The result reads as follows:  
\begin{align}\label{btension}
S^{\rm eff}_{\text{tot}}[g,\gamma]=& \frac{V_F}{V_g}{M_{\rm Pl}^2}\int d^4x\sqrt{g}\left[ R(g)+2m^2\mathcal{U}\left( \sqrt{g^{-1}\gamma_{ab}(\varphi(x))\partial_{\mu}\varphi^a \partial_{\nu}\varphi^b} \right) \right] \nonumber \\
&-\frac{1}{2}\int d^4\varrho d^4\varsigma~h^0_{ab}(\varrho)h^0_{cd}(\varsigma)\Expect{\T^{ab}(\varrho)\T^{cd}(\varsigma)}\,.
\end{align}
The potential  ${\cal U}$  appearing in the above action was defined in (\ref {mGRUK0}).  
Both the CFT and the massive gravity sectors essentially stay the same, despite the global modification. The difference is that the 4D cosmological constant, $\Lambda$, is removed from the $g$-metric dynamics.

To make things crystal clear, we explain a subtlety in how (\ref{btension}) is obtained. Upon varying $\mathcal{A}_5$, the $F$ equations of motion are, schematically, $G_{AB}(F)=(\tilde{\Lambda}+\bar{S})F_{AB}$, where $\tilde{\Lambda}$ is a constant and $G_{AB}$ is the Einstein tensor for $F$. Then the $F$ dynamics depends on a functional of $g$. Meanwhile, the $g$ equations of motion are obtained from $\delta \bar{S}=0$, but this contains the fiducial metric, $\gamma$, which is determined by the bulk dynamics. As such, the $F$ and $g$ equations of motion are coupled in a global, unusual way. If one na\"ively treat $\bar{S}$ as part of the bulk CC, one seems capable of reducing $\mathcal{A}_5$ to a boundary action dual to a pure, unbroken CFT corresponding to the second line of (\ref{btension}). But in the process, not only $F$, but $g$ is also integrated out, because the curvature of $F$ depends on the numerical value of $\bar{S}$ and needs the $g$ equations of motion to be used first.
Technically, it is therefore tricky to just integrate out $F$ but to leave $g$ alone, because of the unusual coupling between $F$ and $g$. However, for most of the reasonable $g$ solutions, $\bar{S}=0$, or $\sim m^2$ if self-accelerated. In view that $m/M_{\rm P} \sim 10^{-60}$, one can make the (incredibly good) approximation to ignore $\bar{S}$ in the $F$ equations. Then we can integrate out only $F$, upon which the boundary action is reduced to (\ref{btension}). There, it looks as if the CFT is softly broken by a scale $m$, but we know it is an illusion, because one gets an unbroken CFT upon integrating out $F$ and $g$ together. This just manifests the fact that the magnitude by which CFT is broken in (\ref{btension}) is commensurate with that of the approximation. This approximation is meaningful, because for now we are only interested in the $g$ sector, not the CFT, which has already been studied exhaustively in the literature.

From (\ref{btension}), the relevant part of the action is
\begin{equation}\label{mGRmod}
\frac{V_{F}}{V_g} M_{\rm P}^2\int d^4x \sqrt{g}\left[ R(g)+2m^2\mathcal{U} \left(\sqrt{g^{-1}\gamma} \right) \right],
\end{equation}
with the fiducial metric $\gamma_{\mu \nu}=\left[\eta_{ab}+h^0_{ab}(\varphi(x)) \right]\partial_{\mu}\varphi^a\partial_{\nu}\varphi^b$, determined  by the Dirichlet boundary data $h^0_{ab}$. Setting $M_{\rm P}$ to 1, the $g$-metric Einstein equations can then be written as nine traceless and one trace equations \cite{Gabadadze:2014rwa},
\begin{equation}
R_{\mu \nu}-\frac{1}{4}g_{\mu \nu}R=T_{\mu \nu}-\frac{1}{4}g_{\mu \nu}T,
\end{equation}
\begin{equation}\label{eom22}
R+T= \langle T \rangle -2m^2 \left\langle g^{\mu \nu} \frac{\partial \mathcal{U}}{\partial g^{\mu \nu}} \right\rangle, 
\end{equation}
\begin{equation}
T_{\mu \nu}=\frac{-2}{\sqrt{g}}\frac{\delta }{\delta g^{\mu \nu}}\int d^4x \sqrt{g}m^2\mathcal{U}=m^2\left( g_{\mu \nu}\mathcal{U}-2\frac{\partial \mathcal{U}}{\partial g^{\mu \nu}} \right),
\end{equation}
where $\langle \cdots \rangle$  denotes the spacetime average,
\begin{equation}
\langle \cdots \rangle\equiv\frac{\int d^4x \sqrt{g} (\cdots)}{V_g}.
\end{equation}

Note that on the self-accelerated solutions, $R,T \sim m^2$ are equal to constants, thus $R=\langle R \rangle$ and $T=\langle T \rangle$ should be used in (\ref{eom22}). Also, $g^{\mu \nu} \partial \mathcal{U} / \partial g^{\mu \nu}=C(\alpha_3,\alpha_4)$ is a constant on the self-accelerated solutions, with $\alpha_3$ and $\alpha_4$ being the free parameters in (\ref{mGRUK0}). Its spacetime average yields the same constant, and therefore (\ref{eom22}) is reduced to
\begin{equation}
R=-2m^2C(\alpha_3, \alpha_4).
\end{equation}  
Hence, for some reasonable choices of parameters, we may get $m^2\sim H_0^2$ in a technically natural way, with $H_0$ being the Hubble constant.\footnote{Note that in massive gravity one would have gotten a similar equation, $ R=-2m^2{\tilde C}(\alpha_3, \alpha_4) $, but with a different function of the parameters, $\tilde C$;
this is because  equation (\ref {eom22}) differs from the corresponding massive gravity equation, $R+T=0$. How this 
difference affects the faith of fluctuations on the self-accelerated solution needs to be investigated.} This conclusion is not changed by introducing quasidilaton \cite{quasidila}, which only affects and improves the dynamics of small perturbations on the solution \cite{Gabadadze:2014kaa}. Furthermore, instead of massive gravity or quasidilaton, one can 
straightforwardly use any of their known extensions. The key point is 
for the theory to have a fiducial metric and for the dark energy to be given by the stress-tensor associated with 
graviton mass and its extension.  As such, our approach establishes a theory on the AdS$_5$ boundary that removes the big cosmological constant and generates a small curvature given by the graviton mass. Meanwhile, it also ascribes natural meanings to the fiducial metric and the diffeomorphism St\"ukelberg fields in the massive gravity.

\section{On Strong Coupling in Massive Gravity}\label{secstrong}

The diffeomorphism invariant action for massive gravity was presented  
in (\ref {mg1}).  The specific structure of the potential guarantees that the theory propagates 
only 5 degrees of freedom.  
It is not guaranteed,  however,  that the above structure is preserved by loop 
corrections. We note  that the coefficients 
$\alpha_{3,4}$ in (\ref {mg1}) get renormalized only multiplicatively \cite {dRGHP}, i.e.  if set to zero they 
remain zero, but the loops would  in general induce other terms such as 
$\det_3(\mathcal{K})^2$ or  $\det_3(\mathcal{K}) \times \det_4(\mathcal{K})$, etc., that 
would reintroduce the sixth ghostly degree of freedom at a certain energy 
scale.

To see this  more explicitly, we take the so-called decoupling limit
$M_{\rm P} \to \infty$, $m\to 0$, with $\Lambda_3 = (M_{\rm P} m^2)^{1/3}$ fixed. In this limit, the  five 
polarizations of the massive graviton  acquire  their individual  identities as the helicity 
$\pm 2$, helicity $\pm 1$ and helicity 0 states. For simplicity, we focus on the helicity-0 
state, which we denote by $\pi$.  The massive gravity  action  in the decoupling limit
contains the following terms for the $\pi$ field:
\be
S_\pi = \int d^4 x  \left ( -{1\over 2} (\d \pi)^2 + \alpha {(\d \pi)^2 \square \pi \over \Lambda_3^3} + 
{2 \alpha^2\over 3}{(\d \pi)^2  
((\d \d \pi)^2-(\square \pi)^2 ) \over \Lambda_3^6} +... \right )\,.
\label{GalS}
\ee
These are the so-called Galileon terms \cite {galileon}; they are special since they 
don't generate higher than two derivatives in the equation of motion, thus 
retaining only one degree of freedom in the $\pi$ theory 
on an arbitrary background.  

On the other hand, it is evident from the above action that interactions of the $\pi$  field 
are described by irrelevant operators, that  become strong  at 
the scale $\Lambda_3$. If this theory  is quantized in a conventional way, the renormalized Lagrangian 
would include an infinite number of counterterms with higher derivatives, such as higher powers 
of $\partial\partial \pi$. 

These new terms, if present, would introduce a ghost at the scale $\Lambda_3$.
One  way to deal with this  problem  is  to regard the action (\ref {GalS})
as an  effective action valid below the scale $\Lambda_3$, above which it needs 
to be completed by some unknown new physics.  The effects of the new physics in the low energy theory
would manifest themselves as an infinite series
of higher derivative operators  suppressed by  $\Lambda_3$. 
The appearance of the ghost at that scale  could then  be attributed  to  the artificial 
truncation of the series at  a finite order.  Such an approach is 
possible, and has so far been often adopted. It  calls for an answer to the  question 
as to what is the completion of massive gravity at the 
scale $\Lambda_3$,  especially given that the latter  is so 
much smaller than $M_{\rm P}$.\footnote{See, for example, \cite{mgreview1, mgreview2},
and the references therein.}

In our approach, however,  such a question does not arise. This is because  the massive 
gravity Lagrangian is not quantized  with $\hbar$,  but instead with $\hbar q$.  The appropriate part of the 
path integral reads schematically as
\be \label{GalZ}
{\rm exp} \left (  {i\over \hbar q}  S_\pi \right )\,.
\ee
Thus, every $\pi$  propagator will be proportional to $\hbar q$ and every 
$\pi$ vertex, be it cubic or quartic, to  $1/ \hbar q$.  If we normalize one-particle states 
in a way that the standard loop expansion is an expansion in powers of the respective 
Planck constant, then all the loop-generated counterterms  will be proportional to 
positive powers of $\hbar q$, and would vanish in the setup considered here.

We note that this does not obviate the strong coupling scale $\Lambda_3$, 
but only renders the full quantum theory with  a finite number of tree-level 
terms, so the theory is still strongly coupled.   While there are no quantum mechanical constraints 
imposed on the amplitudes,  since $\hbar q \to 0$, a resummation of the  classical nonlinear
diagrams would be needed to account for the Vainshtein effect \cite {Arkady}
(see also \cite {DDGV}).  It is also  relevant to note that 
the   tree amplitudes define  loop level amplitudes via the optical theorem which remains valid  order-by-order,  
but  the effective action receives no loop-generated counterterms.

To see how this works in more detail, consider the  $2\to 2$ scattering amplitude at the one-loop level 
that is due to the Wick contraction of two quartic  Galileon terms.   We recall that every  propagator will
 be proportional to $\hbar q$,  and every tree level vertex  will be proportional to   
 $1/ \hbar q$.   Moreover, if we normalize one-particle states 
in a way that  the external lines in Feynman diagrams carry no factors of $\hbar q$,  
then, the one loop diagram for  the  $2\to 2$ scattering amplitude  that comes from a 
contraction of two quartic Galileons would be ${\cal O}(1)$. The imaginary part of this diagram is 
also of order ${\cal O}(1)$, and this is equated  to a square of the tree-level diagram, times
two powers of $\hbar q$ coming from the insertion of the intermediate states, ensuring
the validity of the optical theorem.    While the imaginary part of 
this one-loop diagram would satisfy the optical theorem, the diagram itself  would not give rise 
to a counterterm in the  exponent of the path integral,  since this  counterterm   would  be   
of order ${\cal O}(1)$, while the  existing tree-level terms   are of order $1/\hbar q$.  As to the 
$2\to 2$  amplitude itself, it also receives a  dominant tree-level contribution from the  contraction of two 
qubic Galileon terms, and the corresponding amplitude -- with the normalization chosen here  for  
the external state --  is  of order,  $\hbar q \times (1/\hbar q)^2 =1/\hbar q$.

Thus,  the full quantum effective action has  as  few terms as the 
classical action. All the calculations done in this theory --  although nonperturbative -- 
are  exact.  In other words,  all the Feynman diagrams are defined  just by the tree-level  
action. When external classical sources are present, the Vainshtein mechanism  \cite {Arkady}
(see \cite {Babichev} for a review) will postpone the strongly coupled regime 
to  energy/momentum scales higher than $\Lambda_3$.   

We can then consider the path integral for gravity,
\be
Z_g= {\rm const} \times \int  d\mu(g) d \mu(F) d\lambda\, dq\,  {\rm exp} \left \{{i \over \hbar} \left ( 
\frac{1}{q}(S_{\rm mGR}+\Gamma_{\rm eff})+ \lambda((V_g/V_F)-q) + S_F \right ) \right \}\,,
\label{Zg}
\ee
where $S_{\rm mGR}$ is the massive gravity action for the $g$-universe,  the two measures 
$ d \mu(g) $ and  $ d \mu(F)$  include the gauge fixing conditions and 
the Faddeev-Popov determinants for both $g$ and $F$, and $\Gamma_{\rm eff}$ 
is the 1PI effective action  for the SM fields. Just to reiterate, the SM loops are done in a conventional way using $\hbar$ 
while treating  $g$ and $F$ as  external classical fields, which then gives rise  to  $\Gamma_{\rm eff}$. 
Then, the gravity loops are done using (\ref {Zg}). Thus, the   parameter  $\hbar q$ 
may be regarded as a second  Planck constant that governs the gravity loops.

The mixed gravity-SM loops are done  by integrating  in (\ref {Zg}) with respect to  the SM ``classical 
fields" upon which $\Gamma_{\rm eff}$ depends.  As discussed in Chapter \ref{ch2}, this integration   
leads to a ``double"  quantization of the SM fields  -- the primary quantization  of the SM fields 
being done with $\hbar$, and  the  secondary  quantization  of the ``classical" 
SM fields appearing in the 1PI action  being done with the rescaled 
constant $\hbar q$.  Since in our case $\hbar q \to 0$,
the secondary  quantization does not modify the results of the primary 
quantization of the SM fields.  Thus,  there will be no counterterms due to gravity loops. 
On  the other hand, the optical theorem is valid 
order by order in $\hbar q$ -- it  relates, e.g.,   tree  
amplitudes to the respective loop amplitudes, even though the loop-generated counterterms  
to the effective action tend to  zero as $\hbar q\to 0$.\footnote{We would like to thank V.P. Nair for useful discussions  of the above issues.} 
For instance, consider a one-loop  diagram that describes a  self-energy correction due to a graviton 
to one of the  ``classical" SM fields appearing in $\Gamma_{\rm eff}$. This diagram has  two vertices, 
each  proportional to  $1/\hbar q$,  and two propagators, each proportional to $\hbar q$, hence the diagram is ${\cal O}(1)$,  if  we choose to normalize the one-particle states so that the 
external lines in the diagram carry no powers of $\hbar q$.   
The imaginary part of this diagram, which is also ${\cal O}(1)$, 
describes a classical process of emission of a graviton  by a SM particle. This amplitude 
is certainly nonzero in the $\hbar q \to 0$ limit. However, the  loop diagram itself 
does not introduce any counterterms in the   partition function --  
being ${\cal O}(1)$, it  is suppressed by $\hbar q$, as compared to the leading tree-level 
terms in the path integral that are  of the order $1/\hbar q$.

With the quantization process defined as above, one needs to worry 
about two types of quantum corrections that may potentially ruin the 
classical solution of the cosmological constant problem in  the $g$-universe:
(A)  The action  (\ref {ac1}) and (\ref {acmain})  appear to suggest that there 
should be novel nonlocal interactions in the $g$-universe, arising due to the 
product of two integrals in  $V_g^{-1}S_{\rm mGR}$; 
 (B) The form of the effective action in ({\ref {Zg})   
 raises the  question of whether new polynomial terms   
 of $\lambda$ and $q$  may be generated by quantum corrections, 
 and whether these new terms can spoil the classical solutions. 
 
To address the point (A), let us decompose the metric   
as a background, $g_b$,  and its fluctuation, schematically $g =g_b +h$. The 
inverse volume factor, $V_g^{-1}$, multiplying the  action $S_{\rm mGR}$ , 
can then  be expanded  as follows: $V_g^{-1} = V_b^{-1} - V_b^{-2} H_h+...$, where $V_b = \int d^4 x\sqrt{g_b}$  and 
$H_h = \int d^4 x\sqrt{g_b} h/2$.  The  term,   $- V_b^{-2} H_hS_{\rm mGR}$, as well as  the other 
terms containing higher powers of $h$, will produce new unconventional interaction vertices 
at both the tree level and the loops.  When sandwiched between various states, 
the tree level terms either give trivially zero, or are suppressed by extra powers of the inverse volume, 
$V_b^{-1}$;  they can also give rise to amplitudes  that do   not 
correspond to any scattering process.\footnote{Similar amplitudes are also present in ordinary theories at higher orders 
in $1/\hbar$. For instance,  in a theory of a massive scalar $\phi$, the tree amplitude obtained by 
sandwiching the operator $\hbar^{-2} \int d^4 x \,m^2 \phi^2(x) \times \int d^4y\, m^2 \phi^2 (y)$ between 
two two-particle states does not correspond to any scattering process.} 
As to the loops arising from these  novel vertices, they 
will be suppressed by powers of  $V_b^{-1}$, in addition to being governed by the vanishing 
effective Planck constant,  $\hbar q$.  

To address the point (B),  let us rewrite the partition function 
in  a slightly different  form:
\be
\int \, d \mu(g) d\mu(F)\, d\lambda \, d p \, {\rm exp}  \left \{{i \over \hbar} \left ( 
\frac{V_F}{p}(S_{\rm mGR}+\Gamma_{\rm eff})+ \lambda(V_g -p) + S_F\right ) \right \}\,, 
\label{p}
\ee
where  we  integrate  with respect to $p$  and $\lambda$. Since  there is no Wick contraction 
between $g$ and $F$, the multiplier  $V_F$ in front of $S_{\rm mGR}+\Gamma_{\rm eff}$
does not produce any new vertices in the $g$-universe. Hence, this multiplier can be regarded as a pure number
from the point of view of the $g$-universe.   The new nonlocal vertices described in question (A)   above
are now encoded in the conventional-looking term $\lambda V_g$.  These  new vertices 
will modify  the calculations, but   the 
modifications  are  straightforward to take into account. To do so, let us further rewrite the partition 
function,
\be
\int \, d \mu(g) d\mu(F)\, d\lambda \, d p \, {\rm exp}  \left \{
\frac{i}{{\tilde \hbar}} \left ( S_{\rm mGR}+ \Gamma_{\rm eff} + {\tilde \lambda}(V_g -p) \right) + {i \over \hbar} S_F \right \}\,, 
\ee
where ${\tilde \hbar} =\hbar p/V_F$,  and  ${\tilde \lambda} = \lambda p/V_F$.
Then, due to the loop corrections,  there will be additional terms proportional 
to positive powers of $\tilde \hbar$, which need to be included in the effective action  
for gravity. The  most dangerous  of these terms would be proportional to 
\be
{\tilde \hbar}^{k+1} {\tilde \lambda}^{l+2}\,, \quad {k,l =0,1,2,3, ...}~\,
\label{indterms}
\ee
However, it is easy to see that on the classical solutions of the original theory, 
all  these terms vanish when  $V_g/V_F\to 0$. 

We end this chapter by  commenting on another type of corrections that should be likely to 
exist in the theory.  The action $S_{\rm mGR}$ contains the Einstein-Hilbert term, but it may also contain 
higher derivative terms, such as  higher powers of curvature invariants,
that  are not necessarily induced by quantum loops. These terms can arise in 
a putative UV completion of  gravity.  In string theory, for instance, these terms 
would be due to the $\alpha^\prime$ corrections. Such terms  will not spoil 
any of our arguments as long as we are considering  energies and momenta 
below the scale by which these higher dimensional operators are  suppressed 
in comparison  with the EH term. Since the latter scale 
should be expected to be of order $M_{\rm QG}$, the  effects  
of these higher dimensional operators are then negligible  for the SM fields at and below 
$M_{\rm SM}$.  On the other hand, the higher dimensional operators will  be relevant 
for physics at very short distances. One can hope that these 
operators will smooth out the short length/time singularities  of 
certain classical solutions,  e.g.,  black holes, or cosmological solutions. Similar considerations apply to  
higher dimensional operators  that are also expected to appear in $S_F$.

\chapter{Nonlocal Galileons and Self-Acceleration in Volume-Normalized Massive Gravity} \label{ch5}
This chapter is based on the works: Gabadadze, G., and Yu, S. (2016). Metamorphosis of the Cosmological Constant and 5D Origin of the Fiducial Metric. {\it Phys. Rev. D}, 94:104059. [arXiv:1510.07943 [hep-th]].\footnote{Copyright (2016) by the American Physical Society, \url{https://doi.org/10.1103/PhysRevD.94.104059}.} Gabadadze, G., and Yu, S. (2017). Nonlocal Galileons and Self-Acceleration.
{\it Phys. Lett. B}, 768:397-403. [arXiv:1608.01060 [hep-th]].\footnote{Published under CC BY 4.0, \url{https://doi.org/10.1016/j.physletb.2017.03.027}.}

\section{Summary of Volume-Normalized Massive Gravity}\label{intro}
Continuing the journey from the last chapter, we here address the fate of dark energy in the 
context of the volume-normalized framework: if a CC is removed entirely from the observed universe, then what is the substance  
that  leads to the accelerated expansion? The present cosmological data favors a ``dark fluid" with the equation of state $p=- \rho$.  Refs.   \cite{Gabadadze:2014rwa,ggsy}
propose to use massive gravity   \cite{deRham:2010kj}, 
which gives rise to dark energy that is not a CC,  but has the equation of state $p=-\rho$  
\cite {dRGHP_sa,Theo,Koyama}.\footnote{One can also use extensions of massive gravity, some of which are also 
known to have  healthy self-accelerated solutions. See Refs.  \cite {Mukohyama,Tanaka} and references therein.}

Our focus will be on massive gravity and  its 
self-accelerated solutions \cite {dRGHP_sa,Theo, Koyama},  
embedded in the volume-normalized framework.  In particular, we will work in the decoupling limit, where calculations 
are simple and transparent. This limit corresponds 
to the approximation that all scales considered are 
smaller than the Hubble scale. 

The requirement that the fluctuations of the self-accelerated solution contain no ghost constrains the 
parameter space of the theory. In pure massive gravity,  
the two free parameters, $a_2$ and $a_3$, 
have to be both nonzero and satisfy certain additional constraints to achieve this \cite {dRGHP_sa}. 
In particular, $a_3$ has to be nonzero and positive. 
For $a_3\neq 0$, the scalar-tensor Lagrangian obtained in the 
decoupling limit of massive gravity cannot be diagonalized 
due to a certain nonlinear term proportional to $a_3$. While this by itself  is not  a problem 
but only a technical issue, once we embed the pure massive gravity into the volume-normalized framework, we find that $a_3=0$ is a perfectly good choice 
for which the self-accelerated solution could be stable, given appropriate values of $a_2$.

More unsettling  is the  issue of vector fluctuations on the 
self-accelerated background  in pure massive gravity, who lose both their  kinetic  and the
gradient quadratic terms.  In a classical theory, the nonlinear 
terms are at least quadratic in the vector field, hence a trivial solution ``vector $=0$" could be enforced by the initial data.   
However, quantum fluctuations would invalidate these classical arguments: if no vector field kinetic term is induced by the loops, then the vector quantum fluctuations would be infinitely strongly coupled.
Alternatively,  the loops may generate  a kinetic term 
for the vector field,  and if it has a right sign, this 
would  help  the problem. But in any event,  no clear 
conclusions on the viability of the self-accelerated  solution can be  reached 
without considering the quantum effects, which by themselves   call for 
an appropriate computational scheme since the theory is strongly coupled.  

As discussed in Chapter \ref{ch4}, once the theory of massive gravity is embedded in the volume-normalized framework, the quantum loop corrections in the particle physics sector are governed
by $\hbar$, while the loops in the gravitational sector 
are governed by an effective Planck's constant, $\hbar q$, $q$ being a functional whose value is determined by the classical equations of motion. 
In Chapter \ref{ch4}, one gets $q \to 0$, so the quantum loop corrections 
in the gravitational sector vanish. Thus, at low energies, the tensor, vector, and scalar modes  of gravity need not be 
quantized. Then the vanishing of the vector field quadratic terms
is not a problem for a self-accelerated solution with healthy scalar and tensor 
perturbations, since the solution ``vector $=0$" can be maintained perpetually 
in the absence of quantum fluctuations. 

We here calculate corrections 
to the decoupling limit Lagrangian of massive gravity due to the nonlocal terms
used to solve the CC problem. We show that the obtained theory with the new nonlocal 
Galileon terms has a self-accelerated solution with healthy scalar and tensor fluctuations  
for $a_3=0$, in which case the tensor and scalar 
nonlinear mixing terms can be diagonalized. A special feature of this background is that there is no 
linear scalar-tensor mixing to begin with, and therefore no vDVZ discontinuity \cite {dRGHP_sa}.  
Thus, in the approximation used, linear massive gravity is indistinguishable 
from linear  general relativity  on a de Sitter background of equal curvature; the  two theories  differ only  by  nonlinear  terms.   The vector field still loses its quadratic terms, 
like it does in pure massive gravity, but this is not an issue since we 
can impose ``vector $=0$" without being driven away by the quantum loops.  We also investigate spherically symmetric solutions 
on the self-accelerated background and show their stability for certain 
values of the parameter $a_2$.

It is worth mentioning that C. Cheung and G. Remmen have derived the constraints on the pure massive gravity parameters imposed by unitarity and analyticity of scattering amplitudes \cite {cheung}. Since our nonlocal interactions leave local physics intact, the same constraints would apply to our theory as well. The range of parameters used for the healthy self-accelerated solutions in this work, $0<a_2<1/6$ and $a_3=0$, lies entirely in the allowed region of the parameter space 
computed in \cite {cheung}.

\section{The Framework}\label{theory}

In the framework 
of Ref. \cite {ggsy}, the 4D physical metric, $g_{\mu \nu}$,  lives on the conformal boundary of a 
5D anti-de Sitter (AdS$_5$) space; the latter is endowed by its own 5D metric $F_{AB}$.
In other words,  a  spacetime boundary  -- or an orbifold plane -- has its own 4D metric  $g$  and   is  placed right at the  conformal boundary of AdS$_5$  
described by the metric $F$.  The metrics $g$ and $F$ need not be related to each other. 
The theory  is described by  the  following nonlocal  action functional 
\begin{equation}\label{pr11}
\mathcal{A}=\frac{V_F}{V_g}S+S_F,
\end{equation}
where $V_F$ and $V_g$ are the respective invariant volumes of the 5D  and 4D spacetimes, 
$S_F$ is the bulk Einstein-Hilbert action with a large negative CC, and $S$ is the action of our world, 
containing $g$-gravity and matter fields. 

The action (\ref{pr11}) exhibits  the following property:    
any CC generated in the 4D $g$-universe  by any source --  classical or quantum -- 
is immaterial for the  $g$-universe. Instead, that CC gets absorbed into $S_F$;  
it adds to, or subtracts from,  the curvature of AdS$_5$. 

One should expect  from  (\ref {pr11}) that the gravity and the  particle physics Lagrangians each require 
special treatment at the quantum level \cite{Gabadadze:2014rwa,ggsy}.  
The upshot  is  that the particle physics quantum effects can be maintained  intact, while the gravity 
loops  of $g$ are suppressed by the effective Planck constant, 
$\hbar V_g/V_F$. The ratio of the volumes is determined by the classical solutions such that $\hbar V_g/V_F\to 0$. Then the gravity loops do not generate 
any counterterms  to the  action  for the metric $g$, and 
at low energies, gravity in our universe becomes a purely classical theory that  couples to quantized particle 
physics; $F$-gravity is also quantized in a conventional way   \cite{Gabadadze:2014rwa,ggsy}.  

Dark energy is then introduced  via massive gravity. The action  $S$ that enters (\ref  {pr11}) is
\begin{equation}\label{a1}
S=\int d^4x~\sqrt{g}\left[M_{\text P}^2R-\frac{M_{\text P}^2m^2}{4} \mathcal{U} \left( \mathcal{K}\right)+2\mathcal{L}_{\rm SM} \right]. 
\end{equation}
Here, the graviton mass and the nonlinear interactions are  described  by the dRGT potential \cite{deRham:2010ik,deRham:2010kj},
\begin{equation}
\mathcal{U}(\mathcal{K})=\det_2(\mathcal{K})+\alpha_3\det_3(\mathcal{K})+\alpha_4\det_4(\mathcal{K}),
\end{equation}
where  the matrix $\mathcal{K}=1-A$, $A$ being defined  as  
$A^{\mu}_{~ \alpha}A^{\alpha}_{ ~\nu}=g^{\mu \alpha}\gamma_{\alpha \nu}$, so that 
$\mathcal{K}=1-\sqrt{g^{-1}\gamma}$. The fixed metric $\gamma$ is known as the fiducial metric, which is generally non-flat \cite{hrm}. 
Its origin is unspecified  in the pure dRGT massive gravity, but in (\ref{pr11}), it is directly related to the pullback of the boundary metric induced by bulk dynamics, and specified by the Dirichlet data that determines the bulk metric and sources the boundary CFT fields \cite{ggsy}. In this chapter, we make the easiest choice, that of trivial boundary data and a Minkowski fiducial metric, $\gamma_{\mu \nu}=\eta_{\mu \nu}$. This is consistent with the AdS/CFT prescription \cite{adscft, adscft2, adscft3} (see also \cite{ads2} for review). Meanwhile, the matter fields are still quantized with $\hbar$, so the matter Lagrangian, $\mathcal{L}_{\rm SM}$, should be regarded as the real part of the full quantum 1PI Lagrangian coupled to classical gravity \cite{ggsy}.

Following the common practice, we study the theory (\ref{a1}) by taking the decoupling limit: $M_{\rm P}\rightarrow \infty, m\rightarrow 0, \Lambda_3=(M_{\rm P}m^2)^{1/3}$ and $T_{\mu \nu} / M_{\rm P}$ held fixed. We calculate the new nonlocal terms and find stable vacuum self-accelerated solutions for the so-called  restricted Galileons \cite{galileon, restgal}, that correspond to a special choice of parameters. We emphasize the difference between the self-accelerated solutions in the new theory and in pure massive gravity; in a certain parameter range, the former is ghost-free, but the latter is not \cite{dRGHP_sa}. We then study Schwarzschild solutions on these self-accelerated backgrounds, showing that  
the helicity-0 mode decouples from the source already in the linearized theory \cite {dRGHP}. 

We adopt the mostly plus metric convention in this work. The Levi-Civit\`a symbol, $\varepsilon_{\mu \nu \alpha \beta}$, is normalized so that $\varepsilon_{0123}=1$, $\varepsilon^{0123}=-1$. The contractions of rank-2 tensors will be often denoted by square brackets, for example, $[\Pi]=\Pi^{\mu }_{~\mu}$, $[\Pi^2]=\Pi^{\mu }_{~\nu}\Pi^{\nu}_{~\mu}$, etc. We will also use the following shorthand notations, $\varepsilon \varepsilon \Pi \Pi=\varepsilon^{\mu \alpha \rho \sigma}\varepsilon^{\nu \beta}_{~~~\rho \sigma}\Pi_{\mu \nu}\Pi_{\alpha \beta}$, $\varepsilon_{\mu}\varepsilon_{\nu}\Pi\Pi=\varepsilon_{\mu}^{~~\gamma \alpha \rho}\varepsilon_{\nu \gamma}^{~~~\beta \sigma}\Pi_{\alpha \beta}\Pi_{\rho \sigma}$, $(B^2)^{\mu}_{~\nu}=B^{\mu}_{~\alpha} B^{\alpha}_{~\nu}$, $\varepsilon\varepsilon B\partial A=\varepsilon_{\mu_1 \mu_2 \alpha \beta}\varepsilon^{\nu_1 \nu_2 \alpha \beta}B^{\mu_1}_{~\nu_1}\partial_{\nu_2}A^{\mu_2}$, and so on. A ratio between spacetime integrals is defined in the following way: first compute the ratio of integrals in a finite spacetime volume, then take the limit when the region of integration tend to infinity in all dimensions.

\section{New Terms in the Decoupling Limit}\label{theo}

To  begin with, let us derive the  the new terms that emerge in the decoupling limit 
of massive gravity  due to the nonlocal interactions. The variation of (\ref{a1}) yields
\begin{equation}
\delta \bar{S}\equiv \delta\left( \frac{S}{V_g}\right)=\frac{1}{V_g}\left(\delta S-\frac{S}{V_g}\delta
V_g \right),
\end{equation}
where $V_g$ is a functional of the helicity-2 field, $h_{\mu \nu}$, but depends neither on the helicity-0 field, $\pi$, nor  on 
the helicity-1 field, $A_{\mu}$. Define $h_{\mu \nu}/M_{\rm P}\equiv g_{\mu \nu}-\eta_{\mu \nu}$, then
\begin{equation}\label{expand0}
V_g=\int d^4x~\left[ 1+\frac{h}{2M_{\text P}}-\frac{1}{4M_{\text P}^2}\left( h^{\mu \nu}h_{\mu \nu}-\frac{1}{2}h^2\right)+\mathcal{O}\left(\frac{1}{M_{\text P}^3} \right) \right],
\end{equation}
and therefore
\begin{gather}
\frac{\delta V_g}{\delta h^{\mu \nu}}=\frac{\eta_{\mu \nu}}{2M_{\text P}}-\frac{1}{4M_{\text P}^2}\left( 2h_{\mu \nu}-h\eta_{\mu \nu}\right)+\mathcal{O}\left(\frac{1}{M_{\text P}^3} \right), \label{expand} \\ \frac{\delta V_g}{\delta \pi}=\frac{\delta V_g}{\delta A^{\mu}}=0.
\end{gather}
The expansion in (\ref{expand}) is governed  by  inverse powers of $M_{\text P}$. An integer $n$ will be used  
to denote a respective term in the expansion. We'd like to determine which terms in the product of $S$ 
and (\ref{expand}) survive the decoupling limit in addition to the  known terms  obtained 
in the decoupling limit of massive gravity   \cite{deRham:2010ik}:

The kinetic term is $-\frac{1}{2}h^{\mu \nu}\mathcal{E}^{\alpha \beta}_{\mu \nu}h_{\alpha \beta}$, where $\mathcal{E}^{\alpha \beta}_{\mu \nu}$ is the Einstein operator. This carries no power of $M_{\text P}$, so it  enters 
the decoupling limit action in a straightforward manner.

A general term in the expansion of the potential looks like 
\be
\begin{aligned}
&M_{\text P}^{2-n_{\pi}-n_{A}-n_{h}-n}m^{2-2n_{\pi}-n_{A}}h^{n_h}(\partial A)^{n_A}(\partial^2\pi)^{n_{\pi}} \,,\\
=&(M_{\text P}m^2)^{1-n_{\pi}-\frac{n_A}{2}}M_{\text P}^{1-\frac{n_A}{2}-n_h-n}h^{n_h}(\partial A)^{n_A}(\partial^2\pi)^{n_{\pi}},
\end{aligned}
\ee
where $n_{\pi},n_A,n_h$ are all non-negative integers, and  $1-\frac{n_A}{2}-n_h-n\leq 0$. For a term to survive the decoupling limit, it has to have,  $n_A=n_h=0,n=1$. The resulting novel terms are  precisely the 
nonlinear terms, made of second derivatives of $\pi$, that  are  total derivatives   \cite{deRham:2010ik}. 
It is then necessary to keep them in the full  Lagrangian as we will do below
\begin{equation}
\mathcal{L}=\mathcal{L}_{\text{MG}}+\frac{1}{m^2}\mathcal{L}^{\text{TD}}_{\text{tot}},
\end{equation}
where
\begin{equation}
\mathcal{L}_{\text{MG}}=-\frac{1}{2}h^{\mu \nu}\mathcal{E}^{\alpha \beta}_{\mu \nu}h_{\alpha \beta}+h^{\mu \nu}\sum_{n=1}^3\frac{a_n}{\Lambda_3^{3n-3}}X^{(n)}_{\mu \nu}+\frac{h^{\mu \nu}T_{\mu \nu}}{M_{\rm P}},
\end{equation}
is the  decoupling limit Lagrangian of massive gravity \cite{deRham:2010ik}, and the new terms read as follows
\begin{equation}
\mathcal{L}^{\text{TD}}_{\text{tot}}=\mathcal{L}^{\text{TD}}_2-\frac{1}{\Lambda_3^3}\left( \frac{2a_2}{3}+\frac{1}{3}\right) \mathcal{L}^{\text{TD}}_3+\frac{1}{\Lambda_3^6}\left(\frac{a_2}{6}-\frac{a_3}{2}+\frac{1}{12}\right)\mathcal{L}^{\text{TD}}_4, 
\end{equation}
As it is customary, we have used the following notations:
\begin{gather}
\Pi_{\mu \nu}=\partial_{\mu}\partial_{\nu}\pi, \quad a_1=-\frac{1}{2},\\
X^{(1)}_{\mu \nu}=\varepsilon_{\mu}\varepsilon_{\nu}\Pi,\quad
X^{(2)}_{\mu \nu}=\varepsilon_{\mu}\varepsilon_{\nu}\Pi\Pi,\quad
X^{(3)}_{\mu \nu}=\varepsilon_{\mu}\varepsilon_{\nu}\Pi\Pi\Pi, \\
\mathcal{L}^{\text{TD}}_2=-\frac{1}{2}\varepsilon\varepsilon\Pi\Pi,\quad
\mathcal{L}^{\text{TD}}_3=-\varepsilon\varepsilon\Pi\Pi\Pi,\quad
\mathcal{L}^{\text{TD}}_4=-\varepsilon\varepsilon\Pi\Pi\Pi\Pi.
\end{gather}
The helicity-2 equations of motion then become
\begin{equation}\label{newh}
-\mathcal{E}^{\alpha \beta}_{\mu \nu}h_{\alpha \beta}+\sum_{n=1}^3\frac{a_n}{\Lambda_3^{3n-3}}X^{(n)}_{\mu \nu}=\eta_{\mu \nu}\Pi_*-\frac{T_{\mu \nu}}{M_{\rm P}},
\end{equation}
where
\begin{equation}
\Pi_*=\frac{1}{2\Lambda^3_3}\frac{\int d^4x~\mathcal{L}^{\text{TD}}_{\text{tot}}}{V_0},\quad  V_0= \int d^4x~1.
\end{equation}
In the latter expression we also assume that the four-volume  is regularized by some infrared spacetime box, and that the regularization is removed after all the other calculations are  done. 
The term denoted by $\Pi_*$  is the  new term. It  distinguishes the above equation from its counterpart 
in pure massive gravity obtained in \cite {deRham:2010ik}.    
Note that $\Pi_*$ depends only on the asymptotic values of $\Pi_{\mu \nu}$ 
and is suppressed by $V_0$; its variation with respect to the dynamical fields is therefore zero.  
The helicity-1 field, $A_{\mu}$, decouples completely from the tensor mode 
but appears in the Lagrangian only at the quadratic and higher orders; thus, we can consistently set it to zero 
for simplicity.  Last but not least, the helicity-0 equation of motion is the same as in pure massive gravity 
\cite{deRham:2010ik},
\begin{equation}\label{newpi}
\partial_{\alpha}\partial_{\beta}h^{\mu \nu}\left( -\frac{1}{2}\varepsilon_{\mu}^{~\alpha} \varepsilon_{\nu ~}^{~\beta}+\frac{2a_2}{\Lambda_3^3}\varepsilon_{\mu}^{~\alpha  } \varepsilon_{\nu}^{~\beta }\Pi+\frac{3a_3}{\Lambda_3^6}\varepsilon_{\mu}^{~\alpha } \varepsilon_{\nu}^{~\beta }\Pi\Pi  \right)=0.
\end{equation}
The obtained equations (\ref{newh}) -- (\ref{newpi}) constitute the field equations of our theory 
with the most  general parameters $a_2, a_3$.

In what follows it will also be useful to cast the decoupling limit action into a more conventional form by using the expansion (\ref{expand0})  for $V_g^{-1}$.  The  new term emerging in the decoupling limit  in addition to the ones 
of pure massive gravity is given by the product of $\sim h/M_{\text{P}}$ with the integral of a 
total derivative.  Thus, the  decoupling limit action  can be written as 
\begin{equation}\label{lmg}
\bar{S}_{\rm DL}=\frac{1}{V_0}\int d^4x~ \bar{\mathcal{L}},\quad \bar{\mathcal{L}}= \mathcal{L}_{\text{MG}}-h\Pi_*.
\end{equation}
This could be used to  expand $\bar{\mathcal{L}}$ to study small fluctuations, just like in a local theory. A quick check shows that the variation of $\bar{S}_{\rm DL}$ with respect to the fields gives back the equations of motion, (\ref{newh}) and (\ref{newpi}), as it should be the case. Note also that the above action is invariant 
under linearized  diff-transformations, $h_{\mu\nu}\to  h_{\mu\nu} + \partial_\mu  \zeta_\nu+  \partial_\nu  \zeta_\mu$.


\section{Self-Acceleration with Restricted Galileons}\label{dd}

\subsection{Self-Accelerated Solutions} \label{sa}

Here, we look for self-accelerated solutions in  the theory described in the last section. 
We focus on the case $a_3=0$,\footnote{Healthy self-accelerated solutions also exist when 
$a_3\neq 0$; we leave them to Appendix \ref{mixed}.} which makes the helicity-2 and -0
modes separable under the local field re-definition,
\be \label{redef}
h_{\mu \nu}=\bar{h}_{\mu \nu}+\pi \eta_{\mu \nu}+\frac{2a_2}{\Lambda_3^3}\partial_{\mu}\pi\partial_{\nu}\pi.
\ee
Without loss of generality, fields in the self-accelerated background can be parametrized by 
\be \label{ansatz}
h_{\mu \nu}=-\frac{1}{2} M_{\rm P} H^2x^2\eta_{\mu \nu},\quad \pi=\frac{1}{2} q\Lambda^3_3x^2.
\ee
The above expression for  $h_{\mu\nu}$ guarantees that the Ricci curvature of the background
is $H^2$; the latter and the parameter $q$  are the unknowns of the system.   Furthermore, 
the helicity-2 and -0 equations reduce to, respectively,
\begin{gather}
\frac{M_{\rm P}H^2}{\Lambda_3^3}=-q+(2a_2+2)q^2-\frac{4}{3}(2a_2+1)q^3+\frac{1}{3}(2a_2+1)q^4, \label{H2} \\
2a_2q-\frac{1}{2}=0,
\end{gather}
and the solution  are $q=\frac{1}{4a_2}$, and 
\begin{equation}\label{pi0}
H^2=\frac{(1-4a_2)^2(1-6a_2)}{768a_2^4}m^2.
\end{equation}
The  latter expression for $H^2$  is positive only for $a_2<1/6$.

Note that in the conventional decoupling limit $m\to 0$, and therefore $H\to 0$  if the 
parameter $a_2$ is held fixed. Thus the limit corresponds to the physical  approximation  
of all the distance/time  scales considered  to be much smaller that the 
Hubble  distance/time  scale set by, $H^{-1}$.

One may also  consider a decoupling limit   when $m =  a_2^2 m_0 \to 0$, 
where $m_0$ is some fixed dimensionful parameter.   In the latter case,  
we get $H^2 =  m_0^2/768$.  This limit corresponds to taking all the  distance/time  
scales  to be much smaller than the inverse graviton mass, $m^{-1}$; the  Hubble scale,
$H^{-1}$, is one of such scales, and is held fixed in the limit.  Note that in this limit the classical 
solution for both the $h_{\mu\nu}$ and $\pi$ fields blow up.

\subsection{Stability and Non-superluminality}\label{sa2}

To check the linear stability of the above obtained solutions, we diagonalize the theory by using (\ref{redef}), whereby the pure massive gravity Lagrangian takes the form $\mathcal{L}_{\rm MG}=\mathcal{L}_{\bar{h}}+\mathcal{L}_{\pi}$, $\mathcal{L}_{\bar{h}}$ being the linearized Einstein-Hilbert terms minimally coupled to $T_{\mu \nu}$, and
\begin{equation}\label{old}
\mathcal{L}_{\pi}=\frac{3}{2}\pi \square \pi+\frac{3a_2}{\Lambda_3^3}(\partial\pi)^2\square\pi+\frac{2a_2^2}{\Lambda_3^6}(\partial\pi)^2\left([\Pi^2]-[\Pi]^2\right)+\frac{\pi T}{M_{\rm P}}+\frac{2a_2}{M_{\rm P}\Lambda_3^3}\partial_{\mu}\pi\partial_{\nu}\pi T^{\mu \nu}.
\end{equation} 
This is similar to the cubic and quartic Galileon theories with special coefficients.\footnote{Thus, the diagonalizable subclass of theories are sometimes called ``restricted Galileons," and the non-diagonalizable ones with $a_3\neq 0$, ``mixed Galileons" \cite{restgal, Berezhiani:2013dca}.} Clearly, the helicity-2 modes are automatically stable and exactly luminal. However, for the helicity-0 field, the new nonlocal  term contributes as well; the total 
effective Lagrangian for $\pi$ is
\begin{equation}
\bar{\mathcal{L}}_{\pi}=\mathcal{L}_{\pi}-\Pi_*\left[ \frac{2a_2}{\Lambda_3^3}(\partial{\pi})^2+4\pi \right].
\end{equation}
While the $\pi$ tadpole here is not itself invariant under the galilean transformation, its variation is cancelled by that of $\bar{h}_{\mu 
\nu}$, since $h_{\mu \nu}$ as a whole does not transform. Now, if $a_2<0$, then the last term in (\ref{old}) would have a negative classical renormalization of the $\pi$ kinetic term. For most of the reasonable sources, this would overshoot the sign of $\pi$ kinetic term, whereas the nonlinear terms in (\ref{old}) will only be subdominant \cite{restgal}. Hence, the physical choice for the sign of $a_2$ is $a_2>0$. Combining this with the constraint from the previous
section we  get, $0< a_2 < 1/6$.

Next, consider small perturbations about  a background solution $\pi_b$, 
$\pi=\pi_b+\phi$. Then the Lagrangian,  up to the second order in $\phi$, reads as 
\begin{equation}\label{newphi}
\bar{\mathcal{L}_{\phi}}=\mathcal{L}_{\phi}-\Pi_* \left[ \frac{2a_2}{\Lambda_3^3}(\partial{\phi})^2 +4 \phi \left( 1- \frac{a_2}{\Lambda_3^3}\square \pi_b \right) \right],
\end{equation}
where
\begin{equation}\label{oldmgr}
\begin{aligned}
\mathcal{L}_{\phi}=&\left\{ \left[  -\frac{3}{2}+\frac{6a_2}{\Lambda_3^3}\square \pi_b+\frac{6a_2^2}{\Lambda_3^6}\left((\partial_{\alpha}\partial_{\beta}\pi_b)^2-( \square \pi_b)^2 \right) \right]\eta_{\mu \nu} \right. \\
& \left.-\frac{6a_2}{\Lambda_3^3}\partial_{\mu}\partial_{\nu}\pi_b+\frac{12a_2^2}{\Lambda_3^6}(-\partial_{\alpha}\partial_{\mu}\pi_b\partial^{\alpha}\partial_{\nu}\pi_b+\square\pi_b\partial_{\mu}\partial_{\nu}\pi_b)\right\}\partial^{\mu}\phi\partial^{\nu}\phi  \\
&+\frac{\phi}{M_{\rm P}}\left( T-\frac{4a_2}{\Lambda_3^3}\partial_{\mu}\partial_{\nu}\pi_b T^{\mu \nu} \right) +\frac{2a_2}{M_{\rm P}\Lambda_3^3}\partial_{\mu}\phi\partial_{\nu}\phi T^{\mu \nu}\,,
\end{aligned}
\end{equation}
contains all the terms obtained in pure massive gravity \cite{galileon, restgal}. 

Plugging in the solutions (\ref{ansatz}) -- (\ref{pi0}), the quadratic Lagrangian for $\phi$ becomes
\begin{equation}\label{newperturblag}
\bar{\mathcal{L}}_{\phi}=-\frac{(1-4a_2)^2(1-6a_2)}{128a_2^3}(\partial\phi)^2  +\frac{2a_2}{M_{\text P}\Lambda_3^3}\partial_{\mu}\phi\partial_{\nu}\phi T^{\mu \nu}.
\end{equation}
Note that the $\phi$ tadpole term vanishes, confirming the kinetic de-mixing between the helicity-2 and -0 fluctuations on self-accelerated backgrounds. This happens because the coefficient of the kinetic mixing term is proportional to the $\pi$ equation of motion  \cite{dRGHP_sa}, and therefore the de-mixing is not affected by any cosmological matter present in  $T_{\mu \nu}$. The expression  (\ref{newperturblag}) shows that the scalar fluctuations on the self-accelerated backgrounds with $0<a_2<1/6$ are safely stable and non-superluminal. This differs significantly  from 
pure massive gravity. Indeed, without the extra contribution from the second term in (\ref{newphi}), the quadratic Lagrangian would have been generated by (\ref{oldmgr}) only  and would have taken the form:
\begin{equation}
\mathcal{L}_{\phi}=\frac{3}{4}(\partial\phi)^2+\frac{2a_2}{M_{\text P}\Lambda_3^3}\partial_{\mu}\phi\partial_{\nu}\phi T^{\mu \nu}.
\end{equation}
The latter describes  a ghost for all self-accelerated solutions, as already seen in the literature \cite{dRGHP_sa}. 
Pure massive gravity requires $a_3\neq 0$ for the scalar perturbation to have a positive kinetic term; 
in the present theory  one can get  away even with $a_3=0$, as described above.

Note that  should $a_2>1/6$ (and $\neq 1/4$), the resulting AdS solution with $H^2<0$ 
would have carried a ghost.

\subsection{Spherically Symmetric Source on Self-Accelerated Background} \label{3.3} 

In this subsection, we study the gravitational field configuration created by a spherically symmetric 
source of finite size $R$ and uniform density and pressure, using the techniques of Ref. \cite {GigaDavid}; 
the source has total mass $M$. We would like to see if a  Schwarzschild-like solution exists on the 
healthy self-accelerated backgrounds found above.

We start with the most general spherically symmetric metric parametrization,
\begin{gather}
h_{00}=a(r),\quad h_{ij}=f(r)\delta_{ij}, \\
\pi=\frac{c}{2}\Lambda_3^3t^2+\pi_0(r).
\end{gather}
By integrating the helicity-2 equations, (\ref{newh}), with the condition  that  the fields  
vanish at the origin, we'll   get 
\begin{gather}
\frac{f'}{r}=-\frac{\Pi_*}{3}-\frac{M(r)}{4\pi M_{\text P}r^3}+\Lambda_3^3(\lambda-2a_2\lambda^2),  \label{he21}  \\
\frac{a'}{r}=\frac{2\Pi_*}{3}-\frac{M(r)}{4\pi M_{\text P}r^3}+\Lambda_3^3 \lambda(1+4ca_2)+c\Lambda_3^3. \label{he22}
\end{gather}
Then,  the helicity-0 equation, (\ref{newpi}),  reduces to the following expression
\begin{equation}\label{mastereq}
\begin{aligned}
&8a_2^2\lambda^3-12a_2(1+2ca_2)\lambda^2+\left[ 3+4a_2\left( 3c+\frac{\Pi_*}{\Lambda_3^3} \right) \right]\lambda -\frac{4\Pi_*}{3\Lambda_3^3}(1+ca_2)-c  \\
&=\left\{ \begin{aligned}
&\left( \frac{r_*}{r} \right)^3 (1+4ca_2), \quad r>R, \\
&\left( \frac{r_*}{R} \right)^3(1+4ca_2), \quad r<R. \\
\end{aligned}
\right.
\end{aligned}
\end{equation}
Above  we have used the notations  
\begin{equation}
\lambda(r)=\frac{\pi_0'(r)}{\Lambda_3^3r}, \quad r_*=\left(\frac{M}{4\pi M_{\text P}^2m^2} \right)^{\frac{1}{3}},
\end{equation}
where $r_*$ is known as the Vainshtein radius. Let us calculate the  expression for the 
nonlocal term on the ansatz at hand,
\begin{equation}\label{pistar}
\Pi_*=\frac{1}{2\Lambda^3_3}\frac{\int d^4x~\mathcal{L}^{\text{TD}}_{\text{tot}}}{V_0} =\Lambda^3_3[-3c\lambda_{\infty}-(1+c)(1+2a_2)\lambda_{\infty}^3+3\lambda_{\infty}(1+c+2ca_2)],
\end{equation}
where $\lambda_{\infty}\equiv \lambda(r \rightarrow \infty)$.

We can see that there exists a ``de-mixing" value,  $c=-1/(4a_2)$, which makes the helicity-0 equation independent of the mass of the source. This defines a class of solutions on which the kinetic mixing between helicity-2 and -0 fluctuations vanishes. In this case, $\lambda$ must be constant on solution and equal to its asymptotic value, $\lambda_{\infty}$. The expression (\ref{mastereq}) then becomes an equation for $\lambda_{\infty}$,
\begin{equation}\label{demixroot}
(\lambda_{\infty}-1)(4a_2\lambda_{\infty}-1)[(8a_2^2+2a_2-1)\lambda_{\infty}^2+(2-4a_2)\lambda_{\infty}-1]=0.
\end{equation}
In particular, the solutions with $\lambda_{\infty}=1/(4a_2)$ corresponds to the self-accelerated backgrounds found in the last subsection.\footnote{The other roots of (\ref{demixroot}) also give valid Schwarzschild solutions, only that they correspond to pressureless, positive-energy cosmological backgrounds. We discuss them in Appendix \ref{cosb}.} 

 Solving for the metric components, we obtain
\begin{equation}
a=\frac{M}{4\pi M_{\text P}r}+r^2\frac{(1-4a_2)^2(1-6a_2)}{768a_2^4}\Lambda_3^3,
\end{equation}
\begin{equation}
f=\frac{M}{4\pi M_{\text P}r}-r^2\frac{(1-4a_2)^2(1-6a_2)}{1536a_2^4}\Lambda_3^3.
\end{equation}
Therefore, the effective pressure and energy density of the $\pi$ fluid are, respectively,\footnote{Note that $p$ and $\rho$ here are not calculated from the canonically defined stress-energy tensor, $T_{\mu \nu}$, but from the linearized Einstein tensor. In view of (\ref{newh}), they receive extra contributions from the Galileon and the global terms as well. This comment applies to all pressures and energy densities calculated in this chapter.}
\begin{equation}
p=\frac{M_{\text P}}{3}\nabla^2(f-a)=-\frac{(1-4a_2)^2(1-6a_2)}{256a_2^4}M_{\text P}\Lambda_3^3,
\end{equation}
\begin{equation}
\rho=-M_{\text P}\nabla^2f=\frac{(1-4a_2)^2(1-6a_2)}{256a_2^4}M_{\text P}\Lambda_3^3,
\end{equation}
with the right dS equation of state $p=-\rho$. The curvature can then be restored,
\begin{equation}
H^2=\frac{\rho}{3M_{\rm P}^2}=\frac{(1-4a_2)^2(1-6a_2)}{768a_2^4}m^2,
\end{equation}
and coincides, as it should,  with  that obtained in a different gauge in Subsection \ref{sa}. 
Owing to the kinetic de-mixing, the stability analysis is identical to that of the source-free 
case in the last subsection, so the obtained  Schwarzschild solutions in the range, $0<a_2<1/6$,  
are stable.

So far, we have only considered a special value of the parameter $c$  that leads to the de-mixing phenomenon. 
We want to see if there exist other self-accelerated backgrounds on which star-like objects can survive. 
A necessary condition for such solutions is that as $r \rightarrow \infty$, the $\pi$ field profile tends to $\pi \rightarrow c\Lambda_3^3(t^2-r^2)/2$; this  implies that $c=-\lambda_{\infty}$. By requiring this in (\ref{mastereq}), we obtain
the following algebraic equation 
\begin{equation}\label{selfa}
4\lambda_{\infty}-8\lambda_{\infty}^2(1+3a_2)+\frac{16}{3}\lambda_{\infty}^3(1+8a_2+6a_2^2)-\frac{4}{3}\lambda_{\infty}^4(1+18a_2+32a_2^2)+\frac{16}{3}\lambda_{\infty}^5a_2(1+2a_2)=0,
\end{equation} 
and look for its roots. These roots are 
\begin{equation}\label{roots}
\lambda_{0,1,2,3,4}=\frac{1}{4a_2},~0,~1,~\frac{3}{2}-\frac{\sqrt{3}}{2}\sqrt{\frac{6a_2-1}{2a_2+1}},~\frac{3}{2}+\frac{\sqrt{3}}{2}\sqrt{\frac{6a_2-1}{2a_2+1}}.
\end{equation}
Among them, $\lambda_{\infty}=\lambda_{0}$ corresponds to the de-mixing case already discussed. As to  
the rest of the roots, the helicity-0 equation does depend on the mass of the source, so if $r_*>R$, there is a Vainshtein region sufficiently close to the course, in which $\lambda \gg 1$. This must be able to extrapolate to $\sim \lambda_{\infty}$ outside the Vainshtein region, so $\lambda_{\infty}$ has to be the largest root of (\ref{mastereq}) at $r\rightarrow \infty$, with $c$ set to $-\lambda_{\infty}$ \cite{galileon}. Given that $a_2>0$, this can never be satisfied when $\lambda_{\infty}=\lambda_{1,3}$, for which there is always a larger root. In particular, the existence of the 
obstruction   for  $\lambda_{\infty}=0$  implies that asymptotic flatness must be given up in the presence of a spherically symmetric source. The other cases, $\lambda_{2,4}$, are able to satisfy the interpolation condition given appropriate values of $a_2$, resulting in stable cosmological solutions that are, however, not self-accelerated. We give explicit forms and stability analyses of them in Appendix \ref{cosb}. To conclude, the dS vacua given by $\lambda_{\infty}=1/(4a_2)$, $0<a_2<1/6$ are the unique,  stable, non-superluminal,  self-accelerated backgrounds that have Schwarzschild solutions exhibiting the de-mixing phenomenon.  

The remaining caveat here is that the vector fluctuation kinetic term vanishes on the self-accelerated solutions found above \cite{dRGHP_sa}. In pure massive gravity, this could be a problem since vector fluctuations can couple to scalars infinitely strongly through quantum loops. In this work, there is no quantum effect in the theory, so we can just require that the vector modes are completely turned off due to some initial conditions.

\chapter{Conclusion and Outlook} \label{ch6}
The cosmological constant problem is a difficult problem in fundamental physics with a very unusual nature. The quantum field theory of particle physics and the cosmological theories based on general relativity are both very successful at a high level of precision in terms of describing observable physics, yet they differ by such a startling amount when predicting the vacuum energy density of our universe. Moreover, the vacuum energy density receives contributions from physics at a diverse variety of scales, and it will be very surprising to see that all these contributions nearly cancel each other to somewhat sixty decimal places to generate the rate of cosmic acceleration we observe. To address this unusual crisis, it then seems that an unconventional approach should be welcomed.

We have so far presented such an unconventional solution to the cosmological constant problem in the framework of biverse volume-normalized gravity. In this framework, the big cosmological constant in our universe is eliminated in a nonlocal way that extends Tseytlin's volume-normalized approach and then modifies the cosmological constant in a different universe by a small amount, assuming our universe is globally connected to the other universe. By adopting a two-step quantization scheme, the low energy validity of this classical picture is established, such that the nongravitational matter interactions are quantized with the usual Planck constant, whereas gravity is quantized with a rescaled Planck constant that vanishes if the ratio of the spacetime volume of our universe to that of the other universe tends to zero. The result is then classical gravity and quantum matter in our universe. The nontrivial change to the existing cosmological models is then the removal of the true vacuum of a scalar potential and the absence of the primordial tensor modes. This implies a zero tensor-to-scalar ratio in primordial spectrum and revives the $\phi^2$ model of inflation.

The complete removal of the hard cosmological constant in our universe then necessitates the reintroduction of a dynamical dark energy component. This can be done by various means, such as the Lorentz invariant condensates of vector fields or derivatively interacting scalar fields. We have explored the one possibility of using massive gravity in the volume-normalized framework, which has the virtues of ascribing an origin to the fiducial metric and removing the quantum strong coupling problem in massive gravity. Some of the interesting modifications to self-acceleration in massive gravity are worked out.

There are several directions one could further explore the biverse volume-normalized gravity. First, the action  (\ref{modaction}) allows for straightforward Euclidean extensions,
since the standard rotation to Euclidean times done in both universes simultaneously amounts to the usual Euclidean transition in the generating functional.  Thus, the solutions that 
have real Euclidean counterparts, say maximally symmetric spaces, can 
be discussed using the Euclidean path integral.

Second, a possible connection could be made  to the RS2 model \cite{Randall:1999vf}, where induced 4D gravity exists on the RS brane (set $z=\epsilon>0$  and impose $Z_2$ across the hypersurface 
in Sections \ref{secm1} and \ref{secm2}). This scenario  would allow the boundary metric $\gamma_{ab}$ to be truly dynamical. For instance, its linearized Einstein kinetic term is given in (\ref{kin1}), from which it is clear that a mass scale proportional to $1/\epsilon$ acts as the effective Planck mass for $\gamma_{ab}$. We can perform a similar analysis along the lines of Section \ref{secm2}. However, the value of the CC would need 
to be fine-tuned in the RS model, in order to fulfill a set of junction conditions that maintain the consistency of the RS2 picture, in which case our solution to the big CC problem would be lost. In retrospect, the fine-tuning that precludes our big CC solution in the RS2 scenario is not required on the conformal boundary of AdS$_5$ because there is no $F$-metric dynamics in the $\epsilon\rightarrow 0$ limit.

Last but not least, it would be interesting to  look into the   vierbein formulation of massive gravity
\cite {RachelKurt, Potential}  in the volume-normalized scenario. Besides the St\"uckelberg fields  $\varphi^a$,  the vierbein 
formulation  requires a two-index field $\lambda^{a}_{\bar a}$ \cite{Potential}. It would be   fascinating   
if one were able to understand the origin of this field from a broader context.


\appendix
\chapter{Appendix} \label{appendix}

In this appendix, we provide a few technical derivations and discussions that help clarify and supplement the points in the main text but do not fit in the flow of argument there.

\section*{Boundary Fields in AdS/CFT}\label{append1}
In this section, we switch to the Euclidean signature, in tandem with the standard language of AdS/CFT. We present how to find the dual boundary CFT action for bulk graviton. As a precursor, we discuss the massless scalar case first, which is simpler but very similar to the graviton case.
\subsection*{Massless Scalar}
The Euclidean AdS$_5$ metric is
\begin{equation}\label{metric1}
ds^2=F_{AB}dx^Adx^B=\frac{l^2}{z^2}(dz^2+d\vec{x}^2).
\end{equation}
We set $l=1$ in the following discussion and only restore it when necessary. The scalar action is
\begin{equation}
S=\frac{1}{2}\int d^4xdz~\sqrt{F}F^{AB}\partial_A\Phi\partial_B\Phi,
\end{equation}
whose variation gives the equation of motion
\begin{equation}
\left(\partial_z^2-\frac{3}{z}\partial_z+\square\right)\Phi(z,x)=0.
\end{equation}
The general solution that decays at the horizon $z=\infty$ is $\Phi\sim z^2K_2(z\sqrt{-\square})\phi_0(x)$. Integrating by parts and using equation of motion,
\begin{equation}\label{A4}
S=\frac{1}{2}\int d^4x\sqrt{\tilde{\gamma}}\Phi\hat{n}\cdot\nabla\Phi,
\end{equation}
where $\tilde{\gamma_{ij}}$ is the induced metric on the boundary and $\hat{n}\cdot\nabla=-z\frac{\partial}{\partial z}.$ To obtain the effective boundary action, we set $z=\epsilon$ and take $\epsilon\rightarrow 0$ in the end. We call the boundary condition $\Phi(\epsilon,x)=\phi_0(x)$, so that
\begin{equation}
\Phi(z,p)=\frac{z^2K_2(pz)}{\epsilon^2 K_2(p\epsilon)}\phi_0(p)
\end{equation}
in the Fourier space.

Plugging this into (\ref{A4}),
\begin{equation}
S=-\frac{1}{2}\epsilon^{-4}\int d^4xd^4y~\phi_0(x)\phi_0(y)\int \frac{d^4p}{(2\pi)^4}e^{ip(x-y)}u\frac{\partial}{\partial u}\ln(u^2K_2(u)),
\end{equation}
where $u=p\epsilon.$ We then expand the integrand in series,
\begin{equation}
u\frac{\partial}{\partial u}\ln(u^2K_2(u)) =-\frac{u^2}{2}+\frac{u^4}{4}\left( \ln 2 -\gamma -\ln u \right)+\cdots .
\end{equation}
Here, $\gamma=0.5772...$ is the Euler-Mascheroni constant. The higher order terms in the series can be ignored because they contain more than four powers of $\epsilon$ and drop out when $\epsilon\rightarrow 0$.
 
After performing the Fourier transform, we have\footnote{The relevant formulae are given in \cite{Hinterbichler:2015pta, Freedman:1991tk}, where the nonlocal expressions are ``differentially regularized" to give well-defined Fourier transform.}
\begin{align}\label{ea1}
S[\phi_0]=&-\frac{\epsilon^{-2}}{4}\int d^4x~\phi_0(x)\square\phi_0(x) \nonumber \\
&+\frac{1}{32\pi^2}\int d^4xd^4y~\phi_0(x)\phi_0(y)\square_x^3\left[ \frac{\ln(|x-y|/\epsilon)}{|x-y|}\right].
\end{align}
The content of the second line can be clarified by introducing an arbitrary and finite mass scale $\mu$,
\begin{align}
\square_x^3\left[ \frac{\ln (|x-y|/\epsilon)}{|x-y|^2}\right]&=\frac{1}{2}\square^3_x\left[ \frac{\ln(|x-y|^2\mu^2)}{|x-y|^2}\right]-\square^3_x\left[ \frac{\ln(\mu \epsilon)}{|x-y|^2}\right] \nonumber \\
&=\frac{1}{2}\square^3_x\left[ \frac{\ln(|x-y|^2\mu^2)}{|x-y|^2}\right]+4\pi^2\ln(\mu \epsilon)\square^2_x\delta(x-y).
\end{align}
The first term is the ``differentially regularized" two-point correlator for the boundary scalar CFT operator, up to a constant multiple:\footnote{See Appendix B of \cite{Hinterbichler:2015pta}. Our formulae are obtained by setting $k=2$ there.}
\begin{align}\label{plugin3}
\frac{\pi^2}{24} \Expect{\O(x)\O(y)}_{\text{reg}}&=-\frac{1}{2^7\cdot 6} \square^3_x\left[ \frac{\ln (|x-y|^2\mu^2)}{|x-y|^2} \right] \\
&= \lim_{\kappa\rightarrow 0} \left\{ \frac{1}{|x-y|^{8+2\kappa}}+\frac{\mu^{2\kappa}}{\kappa}\frac{\pi^2}{2^5 \cdot 6} \square^2_x\delta (x-y) \right\},
\end{align} 
so the well-known $\sim |x-y|^{-8}$ two-point function is recovered \cite{2pointfunc}. The total boundary action is
\begin{equation}
S[\phi_0]=-\frac{1}{2}\int d^4x~\phi_0(x)\left(\frac{\epsilon^{-2}}{2}\square-\frac{\ln (\mu\epsilon)}{4}\square^2 \right)\phi_0(x) -\frac{1}{2}\int d^4xd^4y~\phi_0(x)\phi_0(y) \Expect{\O(x)\O(y)}_{\text{reg}}.
\end{equation}

In the AdS/CFT context, the local terms that diverge in the $\epsilon\rightarrow 0$ limit are unphysical and should be removed through holographic renormalization \cite{hologrenorm}. Since the regularized correlator only differs from the unregularized one by local terms, specific renormalization schemes can be constructed to yield either form of the correlator. Still, we note that the differentially regularized expression of the two-point function emerges automatically in the calculation.

\subsection*{Graviton}
Now we can look at the boundary theory induced by a free graviton from the AdS$_5$ bulk. The complete action is \cite{Liu:1998bu,Mueck:1998ug}
\begin{equation}
S=-\int_{M} d^4xdz~\sqrt{F}(R+12)-2\int_{\partial M}d^4x~\sqrt{\tilde{\gamma}}K+6\int_{\partial M}d^4x~\sqrt{\tilde{\gamma}},
\end{equation}
up to a functional of $\tilde{\gamma_{ij}}.$ We set the Planck scale $M_5$ to unity for convenience, and use $i,j,...=1,2,3,4$ to denote the coordinates transverse to $z$, which are raised by $\tilde{\gamma^{ij}}$ on the boundary.

The graviton is defined perturbatively by $F_{AB}= F_{AB}^{(0)}+h_{AB}$ with background solution $ F_{AB}^{(0)}$. We work in the gauge $h_{0B}=0.$ In the equations of motion, we use $h^i_j=\tilde{\gamma}^{ik}h_{kj}$ as the dynamical field, since it is related to the boundary condition in a simple way, $h^i_j|_{z=\epsilon}\equiv \chi^i_j=h^0_{ab}$, where $h^0_{ab}$ is the Dirichlet boundary data.\footnote{Here, $a,b,... =1,2,3,4$ are raised with $\delta^{ab}$. In the case that the boundary condition is related to the fiducial metric of massive gravity, the fiducial metric is $f_{\mu \nu}=(\eta_{ab}+h^0_{ab})\partial_{\mu}\varphi^a\partial_{\nu}\varphi^b$.} This allows the latin indices to be raised and contracted with $\delta^{ij}$ on the boundary, which we will do from now on. The equations of motion are \cite{Mueck:1998ug,Arutyunov:1998ve}
\begin{equation}\label{eom2}
\partial_z^2h_j^i+\square h^i_j-\frac{3}{z}\partial_zh^i_j-\frac{1}{z}\partial_z h\delta^i_j+\partial^i\partial_jh-\partial^i\partial_lh^l_j-\partial_j\partial^lh^i_l=0,
\end{equation}
with constraints
\begin{equation}
\square h-\partial_i\partial^j h^i_j-\frac{3}{z}\partial_zh=0,
\end{equation}
\begin{equation}
\partial_z(\partial_ih-\partial_jh^j_i)=0.
\end{equation}

(\ref{eom2}) implies that the transverse-traceless component of $h^i_j$ satisfies the free massless scalar equation of motion. We can then write the solution of $h^i_j$ in Fourier components, 
\begin{align}\label{h1}
h^i_j(z,x)=&\int\frac{d^4p}{(2\pi)^4}e^{ipx}\left[ \bar{\hat{h}}^i_j\frac{z^2K_2(pz)}{\epsilon^2K_2(p\epsilon)}-\frac{z^2-\epsilon^2}{6}\left(p^ip_j\hat{h}-\frac{p^ip_jp^kp_l}{p^2}\hat{h}^l_k \right) \right. \nonumber \\
&\left. +\frac{p_jp^l}{p^2}\hat{h}^i_l+\frac{p^ip_l}{p^2}\hat{h}^l_j-\frac{p^ip_jp^kp_l}{p^4}\hat{h}^l_k+\frac{1}{3}\left(\delta^i_j-\frac{p^ip_j}{p^2}\right)\left( \hat{h}-\frac{p^kp_l}{p^2}\hat{h}^l_k \right)  \right],
\end{align}
where $\hat{h}^i_j(p)$ is the Fourier transform of $h^i_j|_{z=\epsilon}=\chi^i_j,$ and
\begin{equation}\label{h2}
\bar{\hat{h}}^i_j=\hat{h}^i_j-\frac{p_jp^l}{p^2}\hat{h}^i_l-\frac{p^ip_l}{p^2}\hat{h}^l_j+\frac{p^ip_jp^kp_l}{p^4}\hat{h}^l_k-\frac{1}{3}\left(\delta^i_j-\frac{p^ip_j}{p^2}\right)\left( \hat{h}-\frac{p^kp_l}{p^2}\hat{h}^l_k \right).
\end{equation}
Using equation of motion and integrating by parts, the action becomes \cite{Liu:1998bu,Mueck:1998ug,Arutyunov:1998ve}
\begin{equation}\label{a2}
S=-\frac{1}{4}\epsilon^{-3}\int d^4x (h^j_i\partial_zh^i_j-h\partial_zh)|_{z=\epsilon}.
\end{equation}
We can substitute (\ref{h1}) and (\ref{h2}) into this expression to obtain an effective action for $\chi^i_j$. The nonlocal terms from this action are worked out in \cite{Liu:1998bu}. Here, we give the local terms and the differentially regularized two-point function. The total effective action is
\begin{equation}
S[\chi]= S_{\partial^2}+S_{\partial^4}+S_{\text{nonlocal}},
\end{equation}
analogous to the massless scalar case. To obtain the results below, 4D integration by parts is used when necessary.

$S_{\partial^2}$ receives contribution from both terms in (\ref{a2}):
\begin{equation}\label{d2}
S_{\partial^2}=\frac{\epsilon^{-2}}{2}\int d^4x~\left[-\frac{1}{4}\chi^j_i\square\chi^i_j+\frac{1}{2}\chi^j_i\partial_j\partial^l\chi^i_l+\frac{1}{4}\chi\square\chi-\frac{1}{2}\chi^j_i\partial^i\partial_j\chi \right],
\end{equation}
which has the structure of the kinetic terms in linearized Einstein gravity, as expected.

Both $S_{\partial^4}$ and $S_{\text{nonlocal}}$ come from the first term in (\ref{a2}) only. Like the massless scalar case, the arbitrary mass scale $\mu$ is introduced to the separate these terms. In particular,
\begin{equation}\label{d4}
S_{\partial^4}=\frac{\ln (\mu \epsilon)}{16}\int d^4x~\chi^j_i D^{ik}_{jl}\chi^l_k,
\end{equation}
where the fully symmetrized differential operator $D^{ik}_{jl}$ is
\begin{align}\label{d41}
D^{ik}_{jl}=&\frac{1}{2}\left(\delta^{i}_l\delta^k_j+\delta^{ik}\delta_{jl}-\frac{2}{3}\delta^i_j\delta^k_l\right)\square^2+\frac{2}{3}\p^i\p_j\p^k\p_l \nonumber \\
&-\frac{1}{2} \left( \delta^i_l\p_j\p^k+\delta^{ik}\p_j\p_l+\delta^k_j\p^i\p_l+\delta_{jl}\p^i\p^k-\frac{2}{3}\delta^k_l\p^i\p_j-\frac{2}{3}\delta^i_j\p^k\p_l \right)   \square. 
\end{align}
Also,
\begin{equation}\label{nl1}
S_{\text{nonlocal}}=\frac{1}{128\pi^2}\int d^4xd^4y~\chi^j_i(x)\chi^l_k(y) \square_xD^{ik}_{jl}(x)\left[ \frac{\ln( |x-y|^2 \mu^2 )}{|x-y|^2}\right].
\end{equation}
We can extract from this the regularized two-point function for the graviton CFT operator, 
\begin{equation}\label{TT}
\Expect{\T^i_j(x)\T^k_l(y)}_{\text{reg}}=-\frac{1}{64\pi^2}\square_xD^{ik}_{jl}(x)\left[ \frac{\ln( |x-y|^2 \mu^2 )}{|x-y|^2}\right].
\end{equation}

We must show that this is consistent with the well-known unregularized expression,
\begin{align}\label{nl2}
\Expect{\T^i_j(x)\T^k_l(y)}=~&\frac{20}{\pi^2}\frac{1}{|x-y|^8} \nonumber \\
&\times \left[\frac{1}{2}J^i_l(x-y)J^k_j(x-y)+\frac{1}{2} J^{ik}(x-y)J_{jl}(x-y)-\frac{1}{4}\delta^i_j\delta^k_l \right],
\end{align}
where $J^i_j(x)=\delta^i_j-2\frac{x^ix_j}{x^2}.$ It is straightforward to do so. From Appendix B of \cite{Hinterbichler:2015pta} (also Appendix A of \cite{Freedman:1991tk}), we have the identities
\begin{equation}
\lim_{\kappa\rightarrow 0} \left\{ \frac{1}{|x-y|^{4+2\kappa}}+\pi^2\frac{\mu^{2\kappa}}{\kappa}\delta (x-y) \right\}=-\frac{1}{4} \square_x\left[ \frac{\ln (|x-y|^2\mu^2)}{|x-y|^2} \right],
\end{equation}
\begin{equation}
\lim_{\kappa\rightarrow 0} \left\{ \frac{1}{|x-y|^{6+2\kappa}}+\frac{\pi^2}{8}\frac{\mu^{2\kappa}}{\kappa}\square_x\delta (x-y) \right\}=-\frac{1}{32} \square^2_x\left[ \frac{\ln (|x-y|^2\mu^2)}{|x-y|^2} \right].
\end{equation}
Plugging the non-divergent nonlocal parts of these and (\ref{plugin3}) (terms with negative powers of $|x-y|$) into (\ref{TT}) gives (\ref{nl2}) after some work. We note that this is a more rigorous calculation than the result in \cite{Mueck:1998ug}, which arrives at the same expression but does not address the subtlety of regularizing Fourier transform near poles.

\section*{A Tale of Two Scalars} \label{2scalar}
Here we use a toy model to illustrate the subtleties of diagonalizing a quantum field and a classical field. Consider the following theory of two coupled scalar fields in a Minkowski space,
\begin{equation}\label{toy1}
S_{\rm toy}=\int d^4x\left[-\frac{1}{2}(\partial \sigma)^2-\frac{1}{2}(\partial \tau)^2 -\frac{1}{2}m^2(\sigma^2+\tau^2)-b^2 \sigma \tau  \right],
\end{equation}
where $b^2 \leq m^2$ is a coupling constant. We demand that $\sigma$ is a quantum field but $\tau$ is a classical field, 
being analogs of  the inflaton and metric perturbations, respectively. The fields have commutators
\begin{equation}\label{comm1}
[\sigma(\vec{x},t),\dot{\sigma}(\vec{y},t)]=i\hbar\delta(\vec{x}-\vec{y}), \quad [\tau(\vec{x},t),\dot{\tau}(\vec{y},t)]= 0,
\end{equation}
and all others are zero. 
The usual procedure of diagonalization implements the field redefinitions
\begin{equation}
\Lambda=\sigma+\tau,\quad \Theta= \sigma-\tau,
\end{equation}
upon which
\begin{equation}
S_{\rm toy}=\int d^4x\left[-\frac{1}{4}(\partial \Lambda)^2-\frac{1}{4}(\partial \Theta)^2-\frac{1}{4}(m^2+b^2)\Lambda^2-\frac{1}{4}(m^2-b^2)\Theta^2  \right].
\end{equation}
Naively, if we were to quantize both $\Lambda$ and $\Theta$ (because they each appear to contain a quantum component), it would seem that we are getting two quantum fields out of one quantized  field and one classical field, which cannot be consistent. 
However, (\ref{comm1}) implies that there are now nontrivial cross terms among the new commutators,
\begin{equation}
[\Lambda(\vec{x},t),\dot{\Theta}(\vec{y},t)]=[\Theta(\vec{x},t),\dot{\Lambda}(\vec{y},t)]=i\hbar\delta(\vec{x}-\vec{y}).
\end{equation}
Therefore, $\Lambda$ and $\Theta$ are not independent quantum degrees of freedom, in the sense that a specific combination of them is classical, in this case $\tau=\frac{\Lambda-\Theta}{2}$. Then the ``undiagonalized" form (\ref{toy1}) is actually more suitable for the computation of quantum correlation functions in the standard language of quantum field theory, with $\tau$ treated as a classical source. In the context of inflationary scalar perturbations above, this means that we need not obtain the diagonalized form (\ref{muk}), but just compute the correlation functions of $\delta \phi$ in any gauge to get the correct power spectra.


\section*{Self-Accelerated Solutions for Mixed Galileons}\label{mixed}
Here, we consider the case $a_3\neq 0$ and look for self-accelerated solutions, in the CC-free theory without and with quasidilaton. We specify the parameter space regions in which these solutions are well-defined and stable.

On the ansatz (\ref{ansatz}), the helicity-2 and -0 equations are reduced to, respectively,
\begin{gather}
\frac{H^2}{m^2}= -q+(2a_2+2)q^2+\left(-\frac{8}{3}a_2+2a_3-\frac{4}{3} \right)q^3+ \left(\frac{2}{3}a_2-2a_3+\frac{1}{3} \right)q^4, \\
-\frac{1}{2}+2a_2q+3a_3q^2=0, \label{e2}
\end{gather}
so the solutions are specified by the two roots of (\ref{e2}).

To study stability, consider small perturbations,
\begin{equation}
h_{\mu \nu}=h^b_{\mu \nu}+\chi_{\mu \nu},\quad \pi=\pi^b+\phi.
\end{equation}
The action for the perturbations is
\be
\bar{S}'=\frac{\int d^4x~(\mathcal{L}'-\chi \Pi_*)}{V_0} ,
\ee 
where \cite{dRGHP_sa}
\be
\begin{aligned}
\mathcal{L}'=&-\frac{1}{2}\chi^{\mu \nu}\mathcal{E}^{\alpha \beta}_{\mu \nu}\chi_{\alpha \beta}+6(a_2+3a_3q)\frac{H^2M_{\rm P}}{\Lambda_3^3}\phi\square\phi \\
&+ \frac{a_2+3a_3q}{\Lambda_3^3}\chi^{\mu \nu}X_{\mu \nu}^{(2)}[\Phi]  -3a_3\frac{H^2M_{\rm P}}{\Lambda_3^6}(\partial \phi)^2\square \phi \\
&+\frac{a_3}{\Lambda_3^6}\chi^{\mu \nu}X_{\mu \nu}^{(3)}[\Phi]+\frac{\chi_{\mu \nu}T^{\mu \nu}}{M_{\rm P}},
\end{aligned}
\ee
with $\Phi_{\mu \nu}=\partial_{\mu}\partial_{\nu}\phi$. The kinetic de-mixing between $\chi$ and $\phi$ is explicit, and only $\chi_{\mu \nu}$ couples to sources $T_{\mu \nu}$. The helicity-0 stability condition is then 
\begin{equation}
a_2+3a_3q>0,
\end{equation}
and the positivity of $H^2$ implies
\begin{equation}\label{hel2}
-q+(2a_2+2)q^2+\left(-\frac{8}{3}a_2+2a_3-\frac{4}{3} \right)q^3+ \left(\frac{2}{3}a_2-2a_3+\frac{1}{3} \right)q^4>0.
\end{equation}
Different from pure massive gravity, both the roots in (\ref{e2}) are permitted in the parameter space, giving stable solutions:\\
(I) $q=-\frac{a_2}{3a_3}-\frac{1}{3\sqrt{2}}\sqrt{\frac{2a_2^2+3a_3}{a_3^2}}$, if $a_3<0, 2a_2^2+3a_3>0$ and $-1+4a_2+12a_2^2+24a_3<0$;\\
(II) $q=-\frac{a_2}{3a_3}+\frac{1}{3\sqrt{2}}\sqrt{\frac{2a_2^2+3a_3}{a_3^2}}$, if $a_3>0$ and $-1+4a_2+12a_2^2+24a_3<0$.\\
On top of this, as noted in Subsection \ref{sa2}, we can require $a_2>0$, so that the coupling to source, $\sim \partial_{\mu}\pi\partial_{\nu}\pi T^{\mu \nu}$ has the right sign to renormalize the $\pi$ kinetic term.

\section*{Schwarzschild Solutions on Cosmological Backgrounds}\label{cosb}

We limit to the theory discussed in Section \ref{dd}.
\subsection*{De-mixing Solutions}
The choice $c=-1/(4a_2)$ leads to kinetic de-mixing between the helicity-2 and -0 modes. To find more de-mixing solutions, the other roots of (\ref{demixroot}) are
\begin{equation}
\lambda_{\infty}=1,~\frac{2a_2-1\pm\sqrt{2a_2(6a_2-1)}}{8a_2^2+2a_2-1}.
\end{equation}
We look at them case by case.

$\lambda_{\infty}=1$: We have 
\begin{equation}
a=f=\frac{M}{4\pi M_{\text P}r}-r^2\frac{(1-4a_2)^2}{24a_2}\Lambda_3^3,\quad p=0, \quad \rho=\frac{(1-4a_2)^2}{4a_2}M_{\text P}\Lambda_3^3,
\end{equation}
\begin{equation}
\bar{\mathcal{L}}_{\phi}=2(1-4a_2)^2\left[(\partial_t\phi)^2-\frac{1}{4}(\partial_j\phi)^2 \right],
\end{equation}
so the scalar fluctuation is stable, propagates at half the speed of light, and has positive energy density if $a_2>0$ and $a_2\neq 1/4$.

$\lambda_{\infty}=\lambda_{\pm}=\frac{2a_2-1\pm\sqrt{2a_2(6a_2-1)}}{8a_2^2+2a_2-1}$: We have
\begin{equation}
a=f=\frac{M}{4\pi M_{\text P}r}-r^2\frac{F_{\mp}}{24a_2}\Lambda_3^3,\quad p=0, \quad \rho=\frac{F_{\mp}}{4a_2}M_{\text P}\Lambda_3^3,
\end{equation}
\begin{equation}
\bar{\mathcal{L}}_{\phi}=2F_{\mp}\left[(\partial_t\phi)^2-\frac{1}{4}(\partial_j\phi)^2 \right],
\end{equation}
where
\begin{equation}
F_{\pm}=\frac{(6a_2-1)(32a_2^3+6a_2-1\pm 8a_2\sqrt{2a_2(6a_2-1)})}{(8a_2^2+2a_2-1)^2} >0.
\end{equation}
Again, the fluctuation is stable, propagates at half the speed of light, and has positive energy density if $a_2>1/6$ and $a_2\neq 1/4$. These solutions have zero pressure and positive energy density for $\pi$ fluid.

\subsection*{Solutions with Nontrivial Vainshtein Region}
With $c=-\lambda_{\infty}$, we look at the possibilities corresponding to $\lambda_{\infty}=\lambda_2,\lambda_4$ in (\ref{roots}),
\begin{equation}
\lambda_2=1,\quad \lambda_4=\frac{3}{2}+\frac{\sqrt{3}}{2}\sqrt{\frac{6a_2-1}{2a_2+1}}.
\end{equation}

$\lambda_{\infty}=\lambda_2$: One needs $a_2> 1/4$ for $\lambda_{\infty}$ to be the largest root. To leading order,\\
Outside the Vainshtein radius,
\begin{equation}
\lambda \simeq \lambda_2,\quad f \simeq \frac{M}{4\pi M_{\text P}r},\quad a \simeq \frac{M}{4\pi M_{\text P}r}+r^2\Lambda_3^3(1-4a_2),
\end{equation}
\begin{equation}
\rho \simeq 0, \quad p \simeq 2(4a_2-1)M_{\text P}\Lambda_3^3>0,
\end{equation}
\begin{equation}
\bar{\mathcal{L}}_{\phi} \simeq -\frac{3}{2}(1-4a_2)^2\partial_{\mu}\phi \partial^{\mu}\phi;  
\end{equation}
Inside the Vainshtein radius,
\begin{equation}
\lambda \simeq \frac{(1-4a_2)^{1/3}}{2a_2^{2/3}} \frac{r_*}{r}+\frac{1-2a_2}{2a_2}+\left[\frac{(1-4a_2)^{2/3}}{4a_2^{4/3}}-\frac{(1-4a_2)^{2/3}}{a_2^{1/3}} \right] \frac{r}{r_*},
\end{equation}
\begin{equation}
a\simeq f \simeq \frac{M}{4\pi M_{\text P}r}+\mathcal{O}(r^2\lambda).
\end{equation}
\begin{align}
\bar{\mathcal{L}}_{\phi} \simeq ~& 3\left[(1-4a_2)^{2/3}a_2^{2/3} \left(\frac{r_*}{r}\right)^2+2(1+(1-4a_2)^{1/3}a_2^{1/3})(1-4a_2)\frac{r_*}{r}\right](\partial_t\phi)^2  \nonumber \\
&-3\left[(1-4a_2)^{2/3}a_2^{2/3} \left(\frac{r_*}{r}\right)^2+(1-4a_2)\frac{r_*}{r}\right](\partial_r\phi)^2 \nonumber \\
&-3 \left[  \frac{1-4a_2}{2}\frac{r_*}{r}-\frac{1}{2}+\frac{(1-4a_2)^{1/3}}{4a_2^{1/3}}- (1-4a_2)^{1/3}a_2^{1/3}+8a_2^2 \right] (\partial_{\Omega}\phi)^2.  
\end{align}

$\lambda_{\infty}=\lambda_4$: One needs $a_2> 1/6$ for $\lambda_{\infty}$ to be the largest root. For notational convenience, define
\begin{equation}
\diamondsuit=36a_2^2+12a_2-3, \quad \heartsuit=\frac{12a_2^2-1+2a_2(2+\sqrt{\diamondsuit})}{1+2a_2}.
\end{equation}
Outside the Vainshtein radius,
\begin{equation}
\lambda \simeq \lambda_4,\quad f \simeq \frac{M}{4\pi M_{\text P}r},\quad a \simeq \frac{M}{4\pi M_{\text P}r}-r^2\Lambda_3^3 \frac{72a_2^2+6a_2-3+(12a_2-1)\sqrt{\diamondsuit}}{2(1+2a_2)}  ,
\end{equation}
\begin{equation}
\rho \simeq 0, \quad p \simeq \frac{72a_2^2+6a_2-3+(12a_2-1)\sqrt{\diamondsuit}}{1+2a_2}  M_{\text P}\Lambda_3^3>0,
\end{equation}
\begin{equation}
\bar{\mathcal{L}}_{\phi} \simeq -\frac{3(6a_2-1)[24a_2^2-1+4a_2(1+\sqrt{\diamondsuit})]}{2(1+2a_2)}\partial_{\mu}\phi \partial^{\mu}\phi;  
\end{equation}
Inside the Vainshtein radius,
\begin{equation}
\lambda \simeq -\frac{1}{2a_2^{2/3}}\heartsuit^{1/3} \frac{r_*}{r}-\frac{6a_2^2-1+a_2(1+\sqrt{\diamondsuit})}{2a_2(1+2a_2)}-\frac{1}{4a_2^{4/3}}\heartsuit^{5/3} \frac{r}{r_*},
\end{equation}
\begin{equation}
a\simeq f \simeq \frac{M}{4\pi M_{\text P}r}+\mathcal{O}(r^2\lambda).
\end{equation}
\begin{align}
\bar{\mathcal{L}}_{\phi} \simeq ~& 3\left[a_2^{2/3}\heartsuit^{2/3} \left(\frac{r_*}{r}\right)^2+2a_2^{1/3}\heartsuit^{4/3}\frac{r_*}{r}\right](\partial_t\phi)^2  \nonumber \\
&-3\left[a_2^{2/3} \heartsuit^{2/3}\left(\frac{r_*}{r}\right)^2 +\frac{144a_2^3-10a_2+1+4a_2(6a_2-1)\sqrt{\diamondsuit}}{2(1+2a_2)} \right](\partial_r\phi)^2 \nonumber \\
&-3\left[ \frac{144a_2^3-10a_2+1+4a_2(6a_2-1)\sqrt{\diamondsuit}}{2(1+2a_2)} \right] (\partial_{\Omega}\phi)^2.  
\end{align}

These solutions are free of ghost, gradient instability or superluminality. Inside the Vainshtein region, the radial fluctuation is slightly subluminal, and the angular one is heavily suppressed, as expected \cite{galileon, restgal}. However, far away from the source, the $\pi$ fluid has zero energy density and positive pressure, which cannot possibly match the current observational data.



\cleardoublepage
\phantomsection
\bibliographystyle{apalike}



\end{document}